\documentclass[prb,aps,twocolumn,superscriptaddress]{revtex4-1}
\usepackage{amsmath}
\usepackage{color}
\usepackage{bbm}
\usepackage{amssymb}
\usepackage{epsfig}
\usepackage{multirow}
\usepackage{amsbsy}
\usepackage{array}
\usepackage{diagbox}
\usepackage{bm}
\usepackage{extarrows}
\usepackage{graphicx}
\usepackage{subfigure}
\usepackage{appendix}
\usepackage{txfonts}
\usepackage{lipsum}
\usepackage{bbding}
\usepackage{pifont}
\usepackage{makecell}
\usepackage[version=4]{mhchem}
\graphicspath{{Figures/}}
\allowdisplaybreaks[4]
\usepackage[colorlinks=true,linkcolor=blue,citecolor=blue,urlcolor=blue,bookmarks=false]{hyperref}

\begin{document}
	\title{Identifying Majorana Zero Modes in Vortex Lattices Using Fano Factor Tomography}
	\author{Jiong Mei}
	\affiliation{Beijing National Laboratory for Condensed Matter Physics and Institute of Physics,
	   Chinese Academy of Sciences, Beijing 100190, China}
    \affiliation{School of Physical Sciences, University of Chinese Academy of Sciences, Beijing 100190, China}

    \author{Kun Jiang}
\email{jiangkun@iphy.ac.cn}
    \affiliation{Beijing National Laboratory for Condensed Matter Physics and Institute of Physics,
	   Chinese Academy of Sciences, Beijing 100190, China}
    \affiliation{School of Physical Sciences, University of Chinese Academy of Sciences, Beijing 100190, China}

    \author{Jiangping Hu}
\email{jphu@iphy.ac.cn}
    \affiliation{Beijing National Laboratory for Condensed Matter Physics and Institute of Physics,
	Chinese Academy of Sciences, Beijing 100190, China}
    \affiliation{Kavli Institute of Theoretical Sciences, University of Chinese Academy of Sciences,
	Beijing, 100190, China}
	 \affiliation{New Cornerstone Science Laboratory, 
	Beijing, 100190, China}
 
	\begin{abstract}
		In this work, we investigate the tunneling characteristics of Majorana zero modes (MZMs) in vortex lattices based on scanning tunneling microscopy measurement. We find that zero bias conductance does not reach the quantized value owing to the coupling between the MZMs. On the contrary, the Fano factor measured in the high voltage regime reflects the local particle-hole asymmetry of the bound states and is insensitive to the energy splitting between them. We propose using spatially resolved Fano factor tomography as a tool to identify the existence of MZMs. In both cases of isolated MZM or MZMs forming bands, there is a spatially resolved Fano factor plateau at one in the vicinity of a vortex core, regardless of the tunneling parameter details, which is in stark contrast to other trivial bound states. These results reveal new   tunneling properties of MZMs in vortex lattices and provide measurement tools for topological quantum devices.
	\end{abstract}
	
	\maketitle
	\section{Introduction}
    As building blocks for topological quantum computation \cite{nayak_RevModPhys.80.1083}, Majorana zero modes  (MZMs) have received substantial attention in the past few decades, owing to their predicted non-Abelian statistics, topological protection, and robustness against environmental noise \cite{kitaev2001unpaired,read2000paired,alicea2012new}.
    Owing to previous tremendous efforts, realizing MZMs in materials systems has gained great success, including superconducting proximitized topological insulators \cite{fu2008superconducting}, 1D spin-orbit coupled superconducting nanowires \cite{lutchyn2010majorana,oreg2010helical,lutchyn2018majorana}, ferromagnetic Yu-Shiba-Rusinov chains\cite{nadj2013proposal,PhysRevB.88.155420}, topological planar Josephson junctions\cite{pientka2017topological,hell2017two,ren2019topological,fornieri2019evidence,PhysRevB.106.L241405,PhysRevLett.131.146601}, especially the connate topological superconductor in iron-based superconductors \cite{hu_review,wang2015topological,wu2016topological,xu2016topological,doi:10.1126/science.aao1797}. 
    Among these platforms, vortex-bound states in iron-based superconductors\cite{wang2015topological,wu2016topological,xu2016topological,doi:10.1126/science.aao1797} have shown to be a particularly promising pathway for implementing and studying MZMs. Recently, a large-scale, ordered MZM lattice has also been achieved in naturally strained LiFeAs, leading a pathway towards tunable and ordered MZM lattices \cite{doi:10.1038/s41586-022-04744-8}. In this work, we apply the current noise Fano factor method to the vortex lattices based on scanning tunneling microscopy (STM) and find the Fano factor tomography providing an efficient way of identifying MZMs.

	This paper is organized as follows. We begin by studying single vortices with different bound states using STM measurements. Our study shows that, despite the widely-used differential conductance measurement, spatially resolved Fano factor tomography is an effective tool for distinguishing MZMs from other trivial bound states. Previous literature\cite{PhysRevB.104.L121406} has noted that the Fano factor measured in the high voltage regime reflects the local particle-hole asymmetry of the bound state wave function. Consequently, the spatially resolved Fano factors near an MZM remain at a value of one due to the local particle-hole symmetry of its wave function. On the contrary, the breaking of this symmetry for other trivial zero-energy bound states results in a strong spatially oscillated Fano factor. We conducted an in-depth study on this phenomenon, and while providing the physical picture behind it, we point out that it is robust to the energy splitting between the bound states.
	
	Theoretically, in the case of an isolated MZM, the tunneling conductance is expected to be quantized at zero energy\cite{PhysRevLett.103.237001,PhysRevB.82.180516}, providing a hallmark for their identification. However, the anticipated quantization at zero energy is often deviated in experimental setups due to finite temperature and the presence of couplings between MZMs\cite{doi:10.1126/science.aax0274}. We proceed with studying MZM lattices and focus on the effects of MZM couplings in vortex lattices at zero temperature to comprehend their tunneling characteristics. Particularly, we investigate how MZM couplings suppress the differential conductance. Contrary to the differential conductance, the spatially resolved Fano factors exhibit lower sensitivity to the energy splittings among the MZMs. Hence, Fano factor tomography can reveal the presence of well-separated MBS in the vortex core region, even in the presence of coupling. Furthermore, we investigate the behavior of Fano factors measured in the high voltage regime when both single-electron tunneling and Andreev reflection processes coexist. Our findings reveal that Fano factor tomography can be effective only when Andreev reflections dominate, which necessitates working within the strong tunneling regime.
	
	\section{Single Vortex}\label{sec:2}
\begin{figure}[!htbp]
		\centering
		\includegraphics[width=0.4\textwidth]{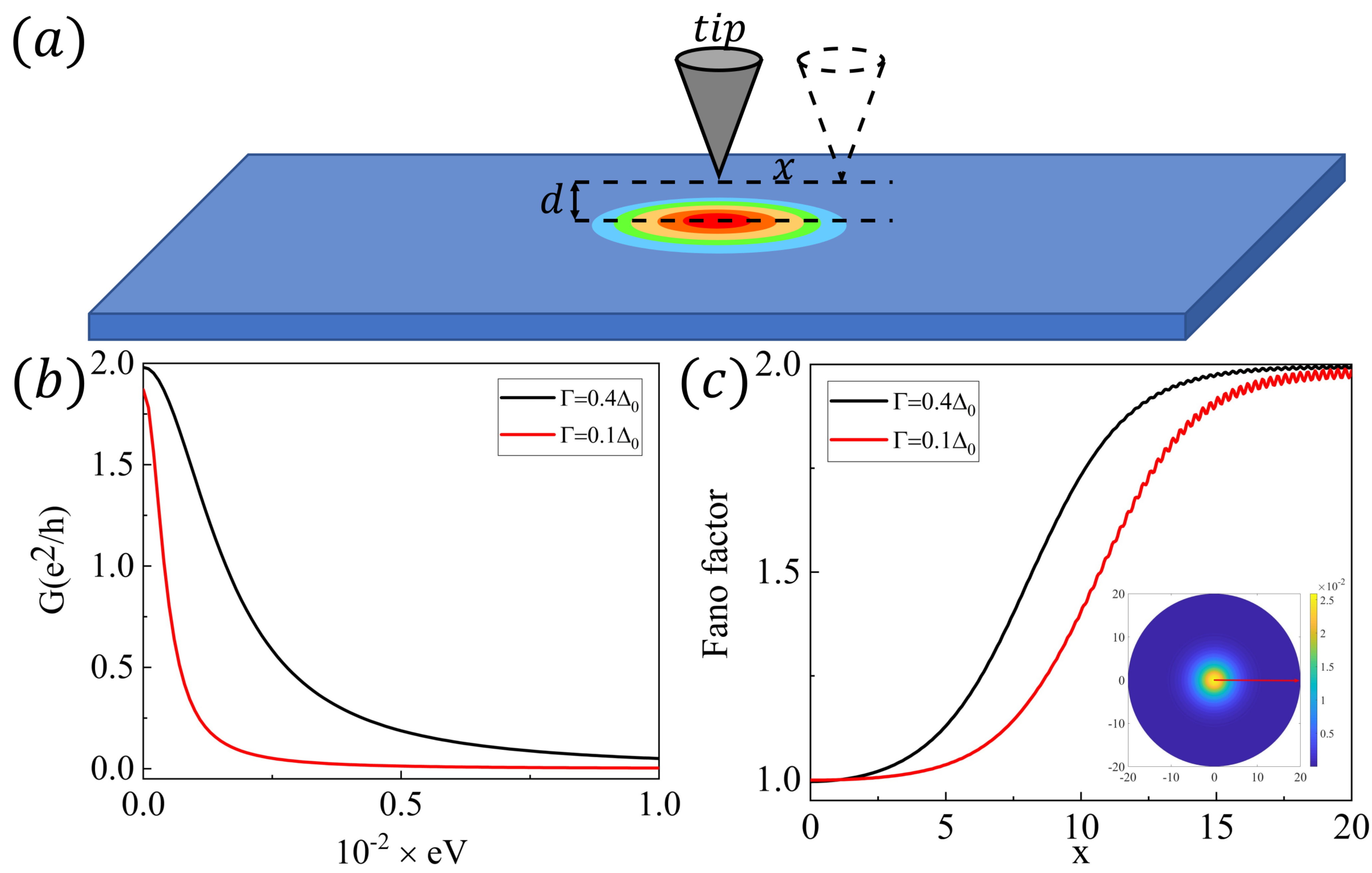}
		\caption{Scheme of a typical STM setup for measuring vortex bound states tunneling characteristics. The energy width $\Gamma$ can be increased by decreasing the tip-sample distance ($d$), and the tip can be moved horizontally to obtain Fano factor tomography. (b) Differential conductance at the vortex core center of a Fu-Kane model. (c) Spatially resolved Fano factors measured in the high voltage regime. The inset in (c) shows the MZM wavefunction, with arrows indicating the scanning directions. The slight oscillations of the Fano factor away from the vortex core are due to the finite bases cutoff in our numerical simulation.}
		\label{fig1}
	\end{figure}

    To gain a complete understanding of the Fano factor tomography, we start with the tunneling characteristics of a single vortex in a 2D superconductor coupled with a metallic STM tip as depicted in Fig. \ref{fig1}(a). To model the STM experiment, we define the complete Hamiltonian $H_{\text{tot}}=H_T + H_S + H_{\text{tunnel}}$, which includes the coupling between the tip and the sample:
	\begin{equation}
		H_{\text{tunnel}} = \sum_{\sigma}t_{\text{tunnel}}\,\psi_{T,\sigma}^{\dagger}c_{j,\sigma}+h.c.
	\end{equation}
	Here, the operator $\psi_{T,\sigma}$ annihilates an electron of spin $\sigma$ at the apex of STM tip while $c_{j,\sigma}$ annihilates an electron of spin $\sigma$ at site $j$ of the 2D lattice. $H_T$ is the Hamiltonian of the isolated metallic tip and $H_S$ is the Hamiltonian of the grounded superconductor. We assume a point contact of the sample-tip tunneling and use a wide-band approximation for the metallic tip. Tunneling events are thus characterized by the energy width $\Gamma = 2\pi\nu_Tt_{\text{tunnel}}^2$, where $\nu_T$ is the density of states in the tip. The bias voltage $V$ between the tip and the sample is taken into account in the chemical potential of the tip as $\mu_T = \mu + eV$. 
	
	The charge current flowing from the tip is $I(j, t_1) = -e\frac{\mathrm{d}\hat{N}_{T}(t_{1})}{\mathrm{d}t_{1}}$, where $\hat{N}_{T}(t_{1})$ is the number operator counting the electrons in the tip at time $t_1$ in the Heisenberg picture. In the DC regime, $I(j, t_1)=I(j)$ and the corresponding differential conductance is defined as $G=\frac{\mathrm{d}I(j,eV)}{\mathrm{d}V}$. Shot-noise is the zero-frequency limit of the time-symmetrized current-current correlator, $S=\int\mathrm{d}(t_1-t_2)S(t_1,t_2)$, given by $S(t_1,t_2)= \left\langle \delta I(t_{1})\delta I(t_{2})\right\rangle +\left\langle \delta I(t_{2})\delta I(t_{1})\right\rangle $ where $\delta I(t_{1})=I(t_{1})-\left\langle I(t_{1})\right\rangle$. Hence, the spatially resolved Fano factor is defined as:
	\begin{equation}\label{sec2:eqn2}
		F(j,eV)\equiv\frac{S(j,eV)}{2e\left|I(j,eV)\right|}.
	\end{equation}
	These physical quantities can be calculated through the standard Keldysh formalism\cite{leeuwen_2013,PhysRevB.104.L121406}. Specifically, when the STM tip tunnels into a pair of single orbital zero energy bound states $\phi_{+}=\left[u_{\uparrow}(j),\,u_{\downarrow}(j),\,v_{\downarrow}(j),\,-v_{\uparrow}(j)\right]^{T}$ and its particle-hole partner, $\phi_{-}=\tau_{y}\sigma_{y}\mathcal{K}\phi_{+}(j)$, where $\mathcal{K}$ denotes the complex conjugation, the Fano factor in the zero-temperature limit and high voltage regime, $eV\gg\Gamma_j$, with $\Gamma_j=\Gamma\sum_{\sigma}(|u_{\sigma}(j)|^2+|v_{\sigma}(j)|^2)$, can have a simple analytical form\cite{PhysRevB.104.L121406}
	\begin{equation}\label{sec2:eqn3}
		F(j)\simeq1+\left(\frac{\sum_{\sigma}(|u_{\sigma}|^{2}-|v_{\sigma}|^{2})}{\sum_{\sigma}(|u_{\sigma}|^{2}+|v_{\sigma}|^{2})}\right)^{2}=1+\delta_{\text{ph}}^{2}(j),
	\end{equation}
	where $\delta_{\text{ph}}(j)$ denotes the local particle-hole asymmetry. Thus, the spatially resolved Fano factors can be used to identify the existence of MZM.
	
	For the 2D superconductor, we use the Fu-Kane model for the superconducting topological surface states \cite{fu2008superconducting} which can be described by the following BdG Hamiltonian,
	\begin{equation}\label{sec2:eqn4}
		H_{S}=\begin{pmatrix}\hat{h} & \hat{\Delta}\\
			\hat{\Delta}^{*} & -\sigma_{y}\hat{h}^{*}\sigma_{y}
		\end{pmatrix},
	\end{equation}
	where $\sigma$ are Pauli matrices acting in the physical spin space, $\hat{\Delta}=\text{diag}(\Delta(\bm{r}),\,\Delta(\bm{r}))$ and $\hat{h}=v_F\bm{p}\cdot\bm{\sigma}-\mu$ represents the surface Hamiltonian of a topological insulator. The Hamiltonian acts on a four-component Nambu spinor $\hat{\Psi}_{\bm{r}}=\left(c_{\uparrow\bm{r}},\,c_{\downarrow\bm{r}},\,c_{\downarrow\bm{r}}^{\dagger},\,-c_{\uparrow\bm{r}}^{\dagger}\right)^{T}$. The SC order parameter in real space is $\Delta(\bm{r})=|\Delta(\bm{r})|e^{i\theta(\bm{r})}$ where $\theta(\bm{r})$ is the SC phase. In the presence of singly quantized vortices located at spatial positions $\left\{\bm{R}_j \right\}$ we may write
	\begin{equation}\label{sec2:eqn5}
		\theta(\bm{r})=\sum_{j}\varphi_{j}(\bm{r}),\quad\varphi_{j}(\bm{r})=\arg(\bm{r}-\bm{R}_{j}).
	\end{equation}
	In addition $\Delta(\bm{r})$ vanishes at the center of each vortex and can be well approximated as $|\Delta(\bm{r})|\simeq\Delta_{0}\prod_{j}\tanh(|\bm{r}-\bm{R}_{j}|/\xi)$. To perform tunneling calculations, we solve this continuous Hamiltonian (\ref{sec2:eqn4}) for the single vortex case directly in a disc with an open boundary condition\cite{Zhu_2016,PhysRevB.43.7609,PhysRevB.82.174506}.

	The differential conductance of a MZM in a vortex core is presented in Fig. \ref{fig1}(b). For this continuous Fu-Kane model, we set $\{v_F,\,\Delta_0,\,\mu\}=\{2.0,\,0.5,\,0.25\}$, leading to a BCS coherence length $\xi=\frac{v_F}{\Delta_0}=4$. In the conductance calculation, we set the inverse temperature as $\beta=\frac{1\times10^4}{\Delta_0}$ and two different energy widths $\Gamma/\Delta_0=0.4,\,0.1$. The quantized zero-bias conductance peak (ZBCP) is observed in the stronger tunneling condition, as shown in Fig. \ref{fig1}(b) with $\Gamma/\Delta_0=0.4$. These stringent requirements (extremely low temperature and strong tunneling condition) make it challenging to achieve the quantized conductance value in experimental setups and also lead to ambiguity in explaining the ZBCP.
	
	Furthermore, based on Eq. (\ref{sec2:eqn3}), Fano factor tomography can be used to detect the presence of MZMs. Fig. \ref{fig1}(c) illustrates the presence of a flat plateau of Fano factors near an isolated MZM (up to $r\sim\xi$), indicating its local particle-hole symmetric nature. Moving away from the core region, the Fano factors increase due to the decreased weight of the wave functions of bound states and the increasing significance of bulk states. In the region sufficiently distant from the vortex core, the dominance of Andreev reflections from the bulk states causes the Fano factor to reach 2. On the contrary, the spatially resolved Fano factors of trivial bound states exhibit strong oscillations between the values of 1 and 2. In the Appendices, we present the simulations for the Caroli-de Gennes-Matricon (CdGM) bound states\cite{CAROLI1964307} and Yu-Shiba-Rusinov (YSR) bound states\cite{Luh1965BOUNDSI,Shiba1968ClassicalSI,Rusinov1969}, similar oscillations of the Fano factors are observed in both cases.

    \section{Fano factors in a Majorana toy model}
\begin{figure}[!htbp]
		\centering
            \subfigure[]{
            \includegraphics[width=0.23\textwidth]{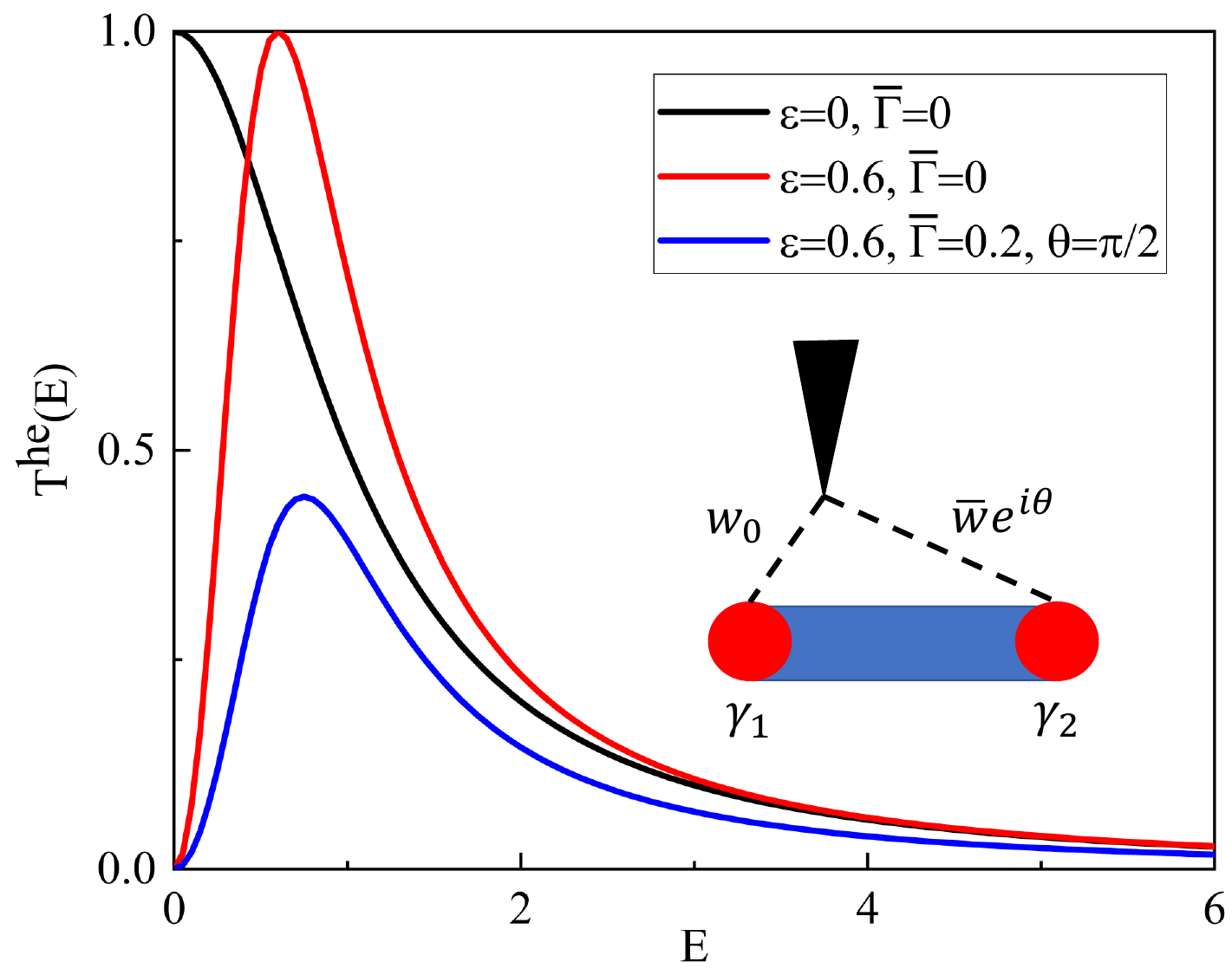}\label{fig2:sub1}
            }
            \subfigure[]{
            \epsfig{figure=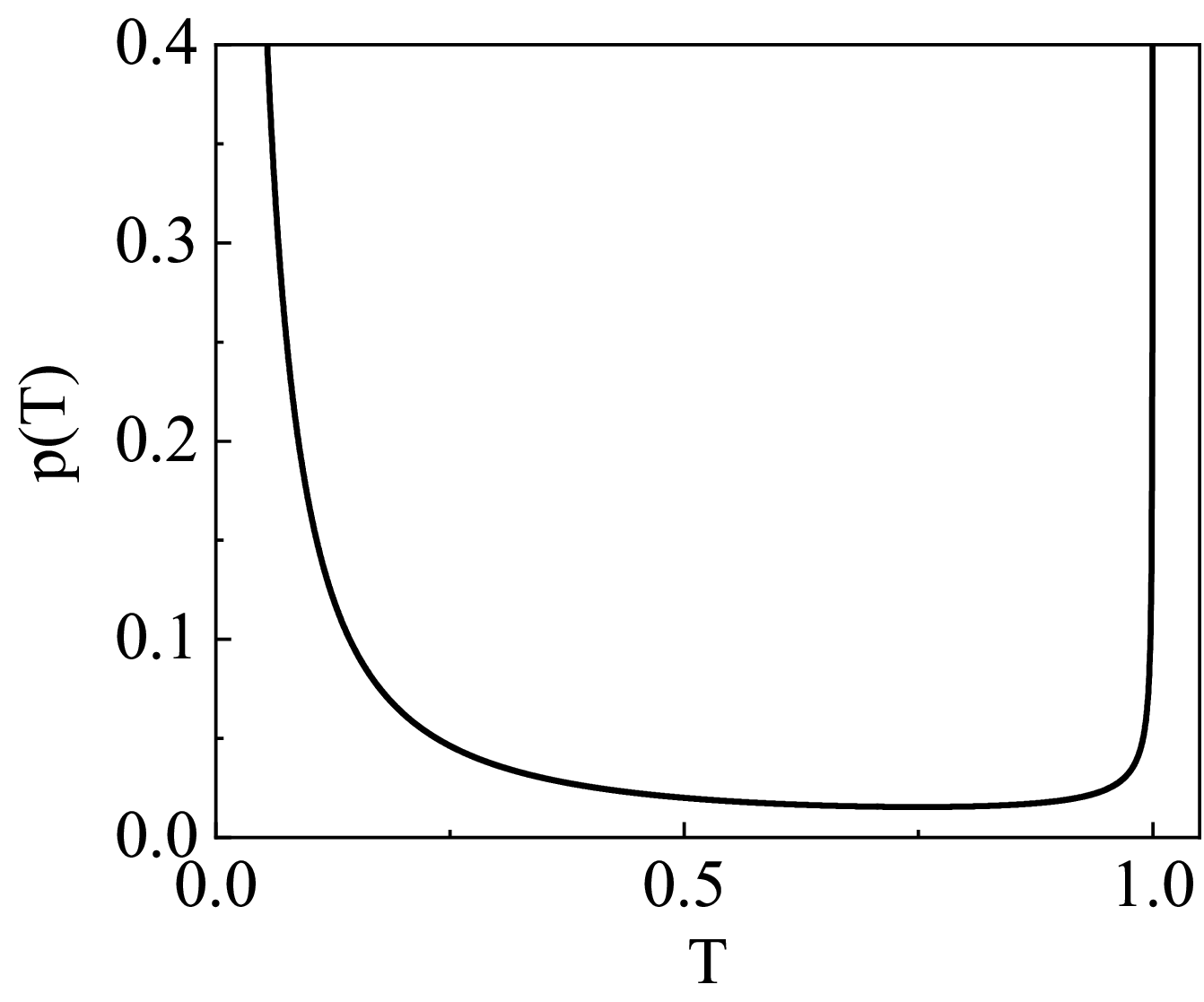, width=0.221\textwidth}\label{fig2:sub2}
            }
		\caption{(a) Schematic plot of the Andreev reflection eigenvalue at different cases where the inset depicts our toy model. Here we set the coupling between the tip and the MBS 1 as $\Gamma_0 = 1.0$. The black line depicts the case of tunneling into an isolated MZM, while in the red and blue lines we turn on the energy splitting between a pair of MBSs. In the blue line, we further turn on the coupling between the tip and MBS 2. (b) The distribution $p(T)$ for the case of tunneling only to one of a pair of Majorana fermions.}\label{fig2}
    \end{figure}

    After gaining the knowledge of a single MZM, we will focus on a pair of Majorana fermions and explore the physical picture behind Eq. (\ref{sec2:eqn3}) in this section. Since Dirac fermions can be expressed as a pair of Majorana fermions, we consider a trivial bound state as a pair of MBSs with finite overlap between their wave functions. Therefore, we model the spatial overlap of the two Majorana wave functions as tunneling into these two MBSs simultaneously, with coupling amplitude $w_0$ and $\bar{w}e^{i\theta}$, respectively. The inset in Fig .\ref{fig2:sub1} depicts our setup. Here, we choose a gauge that the coupling amplitude $w_0$ between the tip and the first MBS is purely real and the coupling with the second MBS can be a generic complex number $\bar{w}e^{i\theta}$. By using scattering formalism\cite{PhysRevLett.101.120403}, the Andreev reflection eigenvalue at energy $E$ can be expressed as\cite{remark1}
        \begin{equation}\label{sec3:eqn1}
		T^{he}(E)=\frac{\left(\Gamma_{0}+\bar{\Gamma}\right)^{2}-4\Gamma_{0}\bar{\Gamma}\sin^{2}\theta}{\left(E-\frac{\varepsilon^{2}+\Gamma_{0}\bar{\Gamma}\sin^{2}\theta}{E}\right)^{2}+\left(\Gamma_{0}+\bar{\Gamma}\right)^{2}},
	\end{equation}
	where $\Gamma_{0}=2\pi w_0^2$, $\bar{\Gamma}=2\pi \bar{w}^2$ and $\varepsilon$ is the energy splitting between these two MBSs. In this single channel tunneling problem, the time averaged current $I$ and shot noise $S$ in the zero-temperature limit, are
    \begin{align}
		I & =\frac{2e}{h}\int_{0}^{eV}\mathrm{d}E\,T^{he}(E)\label{sec3:eqn2}\\
		S & =\frac{8e^{2}}{h}\int_{0}^{eV}\mathrm{d}E\,T^{he}(E)\left(1-T^{he}(E)\right).\label{sec3:eqn3}
    \end{align}
    In the Poisson limit where all the transmission eigenvalue $T^{he}(E)\ll1$, the Fano factor defined in Eq. (\ref{sec2:eqn2}) represents the effective charge of the current carriers which in this case are the Cooper pairs. Under quantum transport conditions, the shot-noise due to the discrete nature of charge is weaker than its classical value since the transmitted electrons are correlated because of the Pauli principle\cite{dejong1996shot,article1997}. From Eq. (\ref{sec3:eqn1}), we can observe that the maximum value of $T^{he}(E)$ is given by
	\begin{equation}
		\max\left(T^{he}(E)\right)=1-\frac{4\Gamma_{0}\bar{\Gamma}\sin^{2}\theta}{\left(\Gamma_{0}+\bar{\Gamma}\right)^{2}},
	\end{equation}
    which can reach the perfect Andreev reflection peak only if $\bar{w}\sin\theta=0$. In this case the coupling matrix between the tip and these two MBSs are purely real, which is equivalent to coupling with only one of the MBSs.

    Obtaining the general expression for the Fano factor with the Andreev reflection eigenvalue (Eq. \ref{sec3:eqn1}) at arbitrary voltage is hard. First, we examine the case of tunneling into an isolated MZM (i.e. $\varepsilon=0,\bar{\Gamma}=0$). In this case, the Andreev reflection eigenvalue simplifies to a Lorentz distribution given by $T^{he}(E)=\Gamma_0^2/(E^2+\Gamma_0^2)$, making it easy to evaluate the integral in Eq. (\ref{sec3:eqn2}, \ref{sec3:eqn3}). Consequently, the Fano factor is\cite{PhysRevB.83.153415,PhysRevB.104.L121406}
	\begin{equation}
		F(V)=1-\frac{eV\Gamma_{0}}{\left[(eV)^{2}+\Gamma_{0}^{2}\right]\arctan\left(\frac{eV}{\Gamma_{0}}\right)}.
	\end{equation}
	In the high voltage limit ($eV\gg\Gamma_{0}$), the Fano factor approaches 1. The suppression of the Fano factor to half the Poisson limit can be attributed to the high transparency peak at $E=0$ in $T^{he}$, as indicated by Eq. (\ref{sec3:eqn3}). When the tip tunnels into both MBSs and the tunneling matrix cannot be gauged into a purely real one, the transparency peak diminishes as $\bar{\Gamma}$ increasing (as long as $\bar{\Gamma}<\Gamma_{0}$), as shown in Fig. \ref{fig2:sub1}. This reduction alleviates the suppression in the quantum transport as described in Eq. (\ref{sec3:eqn3}), subsequently increasing the Fano factor. A finite energy splitting primarily shifts the peak of $T^{he}$ (Eq. (\ref{sec3:eqn1}), and Fig. \ref{fig2:sub1}), but it does not impact the Fano factor in the high voltage regime (details in Appendix C).

    One can transform the uniform distribution in the energy interval $[0,\,eV]$ in the integral (\ref{sec3:eqn2}) and (\ref{sec3:eqn3}) into the distribution of the transmission eigenvalue $T$. It turns out when tunneling into one of the two MBSs, the distribution $p(T)\propto\frac{1}{T^{3/2}\sqrt{1-T}}$ in the limit of $eV\gg\Gamma_i,\varepsilon$, as depicted in Fig. \ref{fig2:sub2}. The suppression of the shot noise below Poisson limit is a consequence of the bimodal distribution of transmission eigenvalues\cite{PhysRevB.46.1889}. Instead of all $Ts$ being close to the average transmission probability, the $Ts$ are either close to 0 or to 1. This reduces the integral of $T(1-T)$. The specific form of this bimodal distribution gives rise to the suppression factor $1/2$. It is worth noted that the suppression 1/2 of the Fano factor not only occurs in tunneling to an isolated MZM but also happens in the symmetric double barrier junctions\cite{PhysRevB.43.4534,dejong1996shot}. In the MZM case, the symmetric tunneling is guaranteed by the particle-hole symmetry of the Majorana wave function.

    The distribution of transmission eigenvalues in a system with many pairs of Majorana fermions would typically deviate from the universal distribution $p(T)$ mentioned earlier. However, as discussed in Appendix D, if the couplings between these Majorana pairs are small compared to the tunneling energy width $\Gamma$, the Fano factor measured from the point contact measurement in the high voltage regime will still be almost 1.
 
	\section{MZMs in vortex lattices}
	Finally, we want to address the MZMs inside vortex lattices. It is well known that certain types of topological superconductors and superconducting topological surface states can host protected MZMs in the cores of Abrikosov vortices\cite{read2000paired,fu2008superconducting}. When these vortices are arranged in a dense periodic lattice, it is expected that the zero modes from neighboring vortices will hybridize and form dispersing bands\cite{PhysRevB.82.094504,PhysRevLett.111.136401,PhysRevB.92.134519}. This section primarily focuses on the square vortex lattice and examines the related properties of differential conductance and Fano factor tomography. The extension to other vortex lattices is straightforward. We discover that the conductance is significantly suppressed due to the couplings between MZMs located at different vortices, while Fano factor tomography can still be used to detect the existence of MZMs.
	
	For a general MZM network, we can write down a general tight-binding Hamiltonian as,
	\begin{equation}
		H_{M}=\frac{i}{2}\sum_{ij}t_{ij}\gamma_{i}\gamma_{j},
	\end{equation}
	where $\gamma_{i}$ is a Majorana operator that satisfies $\gamma_{i}^{\dagger}=\gamma_{i}$ and $\left\{ \gamma_{i},\gamma_{j}\right\} =2\delta_{ij}$. The point contact tunneling Hamiltonian between the normal tip and the SC can be described by:
	\begin{equation}
		H_{T}=\sum_{\sigma}\int\mathrm{d}x\,\left[\tilde{t}\delta(x)\,\psi_{T,\sigma}^{\dagger}(x)\psi_{S,\sigma}(x)+h.c.\right]
	\end{equation}
	When $(eV, \Gamma)\ll\Delta$, the current is dominated by the low-lying Majorana states. Under Nambu representation, $\Psi=\left(\psi_{\uparrow},\,\psi_{\downarrow},\,\psi_{\downarrow}^{\dagger},\,-\psi_{\uparrow}^{\dagger}\right)^{T}$, the projection of the field operator $\Psi$ onto the Majorana states manifold can be approximated as $\Psi(x)\approx\sum_{i}\gamma_{i}\left[f_{\uparrow,i}(x),\,f_{\downarrow,i}(x),\,f_{\downarrow,i}^{*}(x),\,-f_{\uparrow,i}^{*}(x)\right]^{T}$. This approximation leads to the effective tunnel Hamiltonian that describes the coupling between the tip and the Majorana states as,
	\begin{equation}
		H_{T}=\tilde{t}\sum_{\sigma,i}\left(\psi_{T,\sigma}^{\dagger}(0)f_{\sigma,i}(0)-\psi_{T,\sigma}(0)f_{\sigma,i}^{*}(0)\right)\gamma_{i}.
	\end{equation}
    
	In the practical case where the MZMs are well separated and the tip is located at a vortex core center, we can simplify the problem by considering a single Majorana bound state coupled to the tip. This can be achieved by defining the energy width matrix as $\tilde{\Gamma}_{ij}\equiv\delta_{i,0}\delta_{j,0}\tilde{\Gamma}$, where without loss of generality we denote this MBS as the zero-th MBS. Moreover, in the wide band limit, $\tilde{\Gamma}$ is energy independent and $\tilde{\Gamma}=2\pi\nu_T\tilde{t}^2\left(|f_{\uparrow}|^2+|f_{\downarrow}|^2\right)$, where $f_{\sigma}$ is the zero-th Majorana wave function evaluated at the core center. In this scenario, at zero temperature, the differential conductance is given by \cite{PhysRevB.82.180516}
	\begin{equation}\label{sec4:eqn4}
		G(V)=-\frac{2e^{2}}{h}\tilde{\Gamma}\mathrm{Im}[G_{00}^{R}(eV)].
	\end{equation}
	Here, $G_{00}^{R}=g^R_{00}/\left(1+i\tilde{\Gamma}g^R_{00}\right)$ and $g^R(\omega)=2\left[\omega-2i\bm{t}\right]^{-1}$ represents the free retarded Green function of the Majorana network.
	
\begin{figure}[!htbp]
		\centering
		\subfigure[]{
			\includegraphics[width=0.2\textwidth]{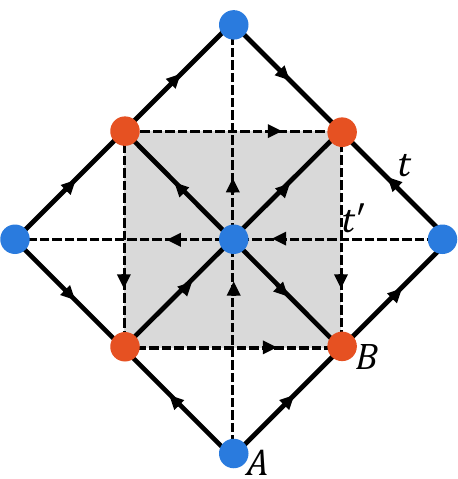} \label{fig:MZM Network Scheme}
		}
		\subfigure[]{
			\epsfig{figure=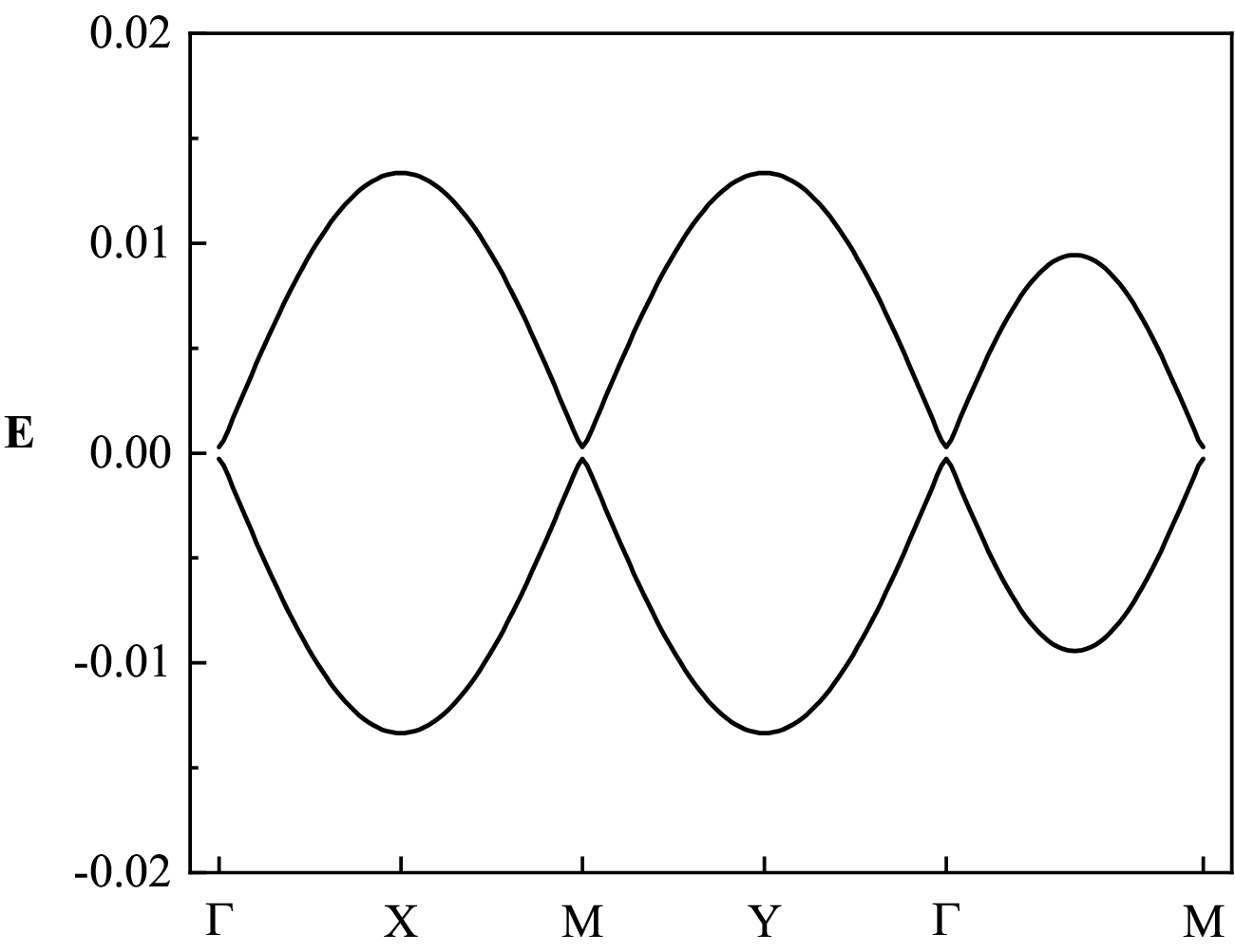, width=0.25\textwidth} \label{fig:MZM TB Bands}
		}
		\caption{(a) A diagrammatic sketch of a square vortex lattice with two vortices in one unit cell (the shaded region). Each vortex core binds a MZM, and the arrows specify the $\mathbb{Z}_2$ gauge factors for the MZM tight-binding models. Hopping in the direction of the arrow incurs a phase factor of $+i$ while hopping in the opposite direction $-i$. (b) A typical band structure when MZMs at different core sites hybridize to form a band. The values of $t$ (NN hopping) and $t'$ (NNN hopping) are $t=2.36\times10^{-3}$  and $t'=3.45\times10^{-5}$, respectively.}
		\label{fig3}
	\end{figure}

\begin{figure}[!htbp]
		\centering
		\epsfig{figure=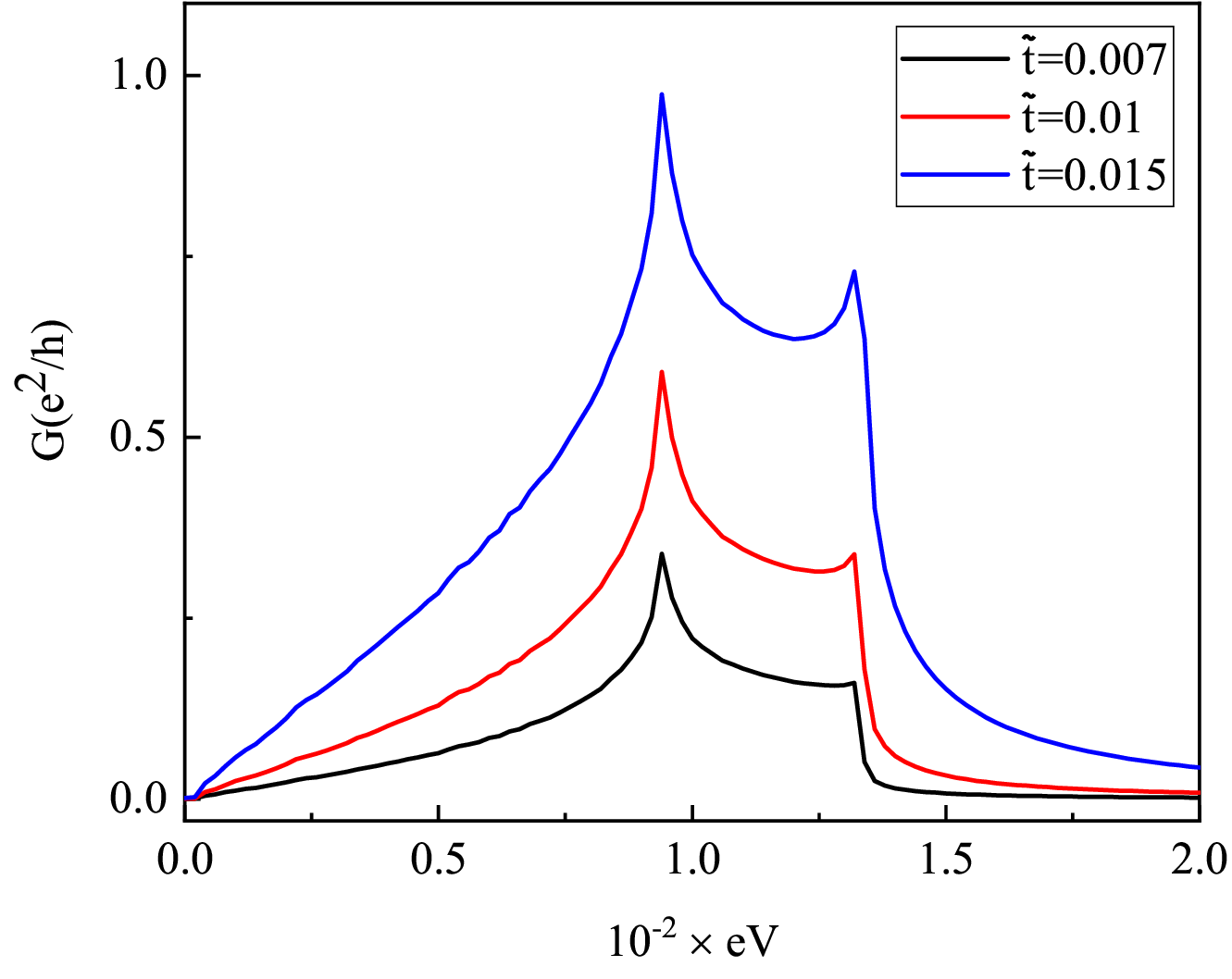, width=0.35\textwidth} \label{fig:G of MZM TB}
		\caption{The tunneling differential conductance of MZM networks at zero temperature. The parameters of the MZM network are the same as in the Fig. \ref{fig:MZM TB Bands}, and we set $\nu_T$ = 0.8, $|f_\uparrow|^2+|f_\downarrow|^2=1$ in the simulations. Under moderate tunneling strength, the conductance peak is significantly suppressed from the ideal quantized value of 2 when the hybridization is strong.}
		\label{fig4}
	\end{figure}

	Now we introduce the tight-binding Hamiltonian of the MZMs in the square vortex lattice. In order to properly account for the gauge field of the system, it is necessary to consider a two-vortex unit cell, as shown in Fig. \ref{fig:MZM Network Scheme}. The Hamiltonian for this system can be written as\cite{PhysRevB.92.134519},
	\begin{align}
		H_{\square}&=\frac{it}{2}\sum_{R}\gamma_{R}^{A}\left(\gamma_{R}^{B}-\gamma_{R-\hat{x}-\hat{y}}^{B}+\gamma_{R-\hat{x}}^{B}+\gamma_{R-\hat{y}}^{B}\right)+h.c.\notag \\ 
		&+\frac{it'}{2}\left[\gamma_{R}^{A}\left(-\gamma_{R+\hat{x}}^{A}+\gamma_{R+\hat{y}}^{A}\right)+\gamma_{R}^{B}\left(\gamma_{R+\hat{x}}^{B}-\gamma_{R+\hat{y}}^{B}\right)\right]+h.c.
	\end{align}
	where the superscripts $A$ and $B$ denote two sublattices in the MZMs network and we have included nearest-neighbour (NN) and next nearest-neighbour (NNN) couplings. Fig. \ref{fig:MZM TB Bands} depicts a typical MZM band structure with a particle-hole symmetric bandwidth around $8\sqrt{2}t$.  Fig. \ref{fig4} displays the corresponding zero-temperature differential conductance when the STM tip is located above one MZM. As shown in Fig. \ref{fig4}, compared to the isolated MZM case where the ZBCP is quantized at value of 2, the peak height of the tunneling differential conductance of an MZM lattice is strongly suppressed to near one due to the coupling between MZMs, which vanishes outside the MZM band. This further complicates the explanation of the ZBCP which casts a shadow over the confirmation of the existence of MZM.
	
		\begin{figure*}[htbp]
		\begin{minipage}[b]{0.95\textwidth}
			\centering
			\subfigure[]{
				\epsfig{figure=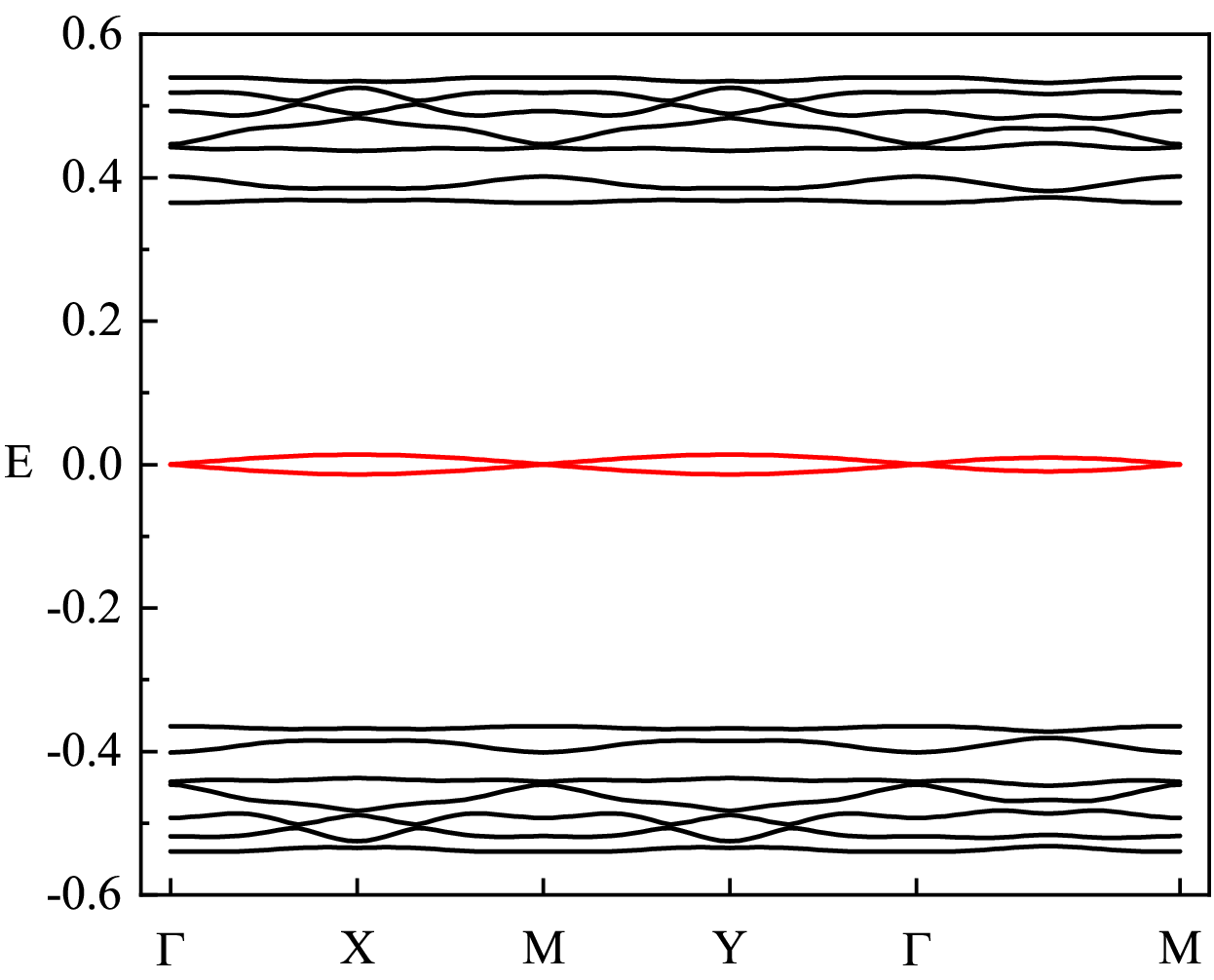, width=0.3\textwidth} \label{fig:Lattice_Bands1}
			}
			\subfigure[]{
				\epsfig{figure=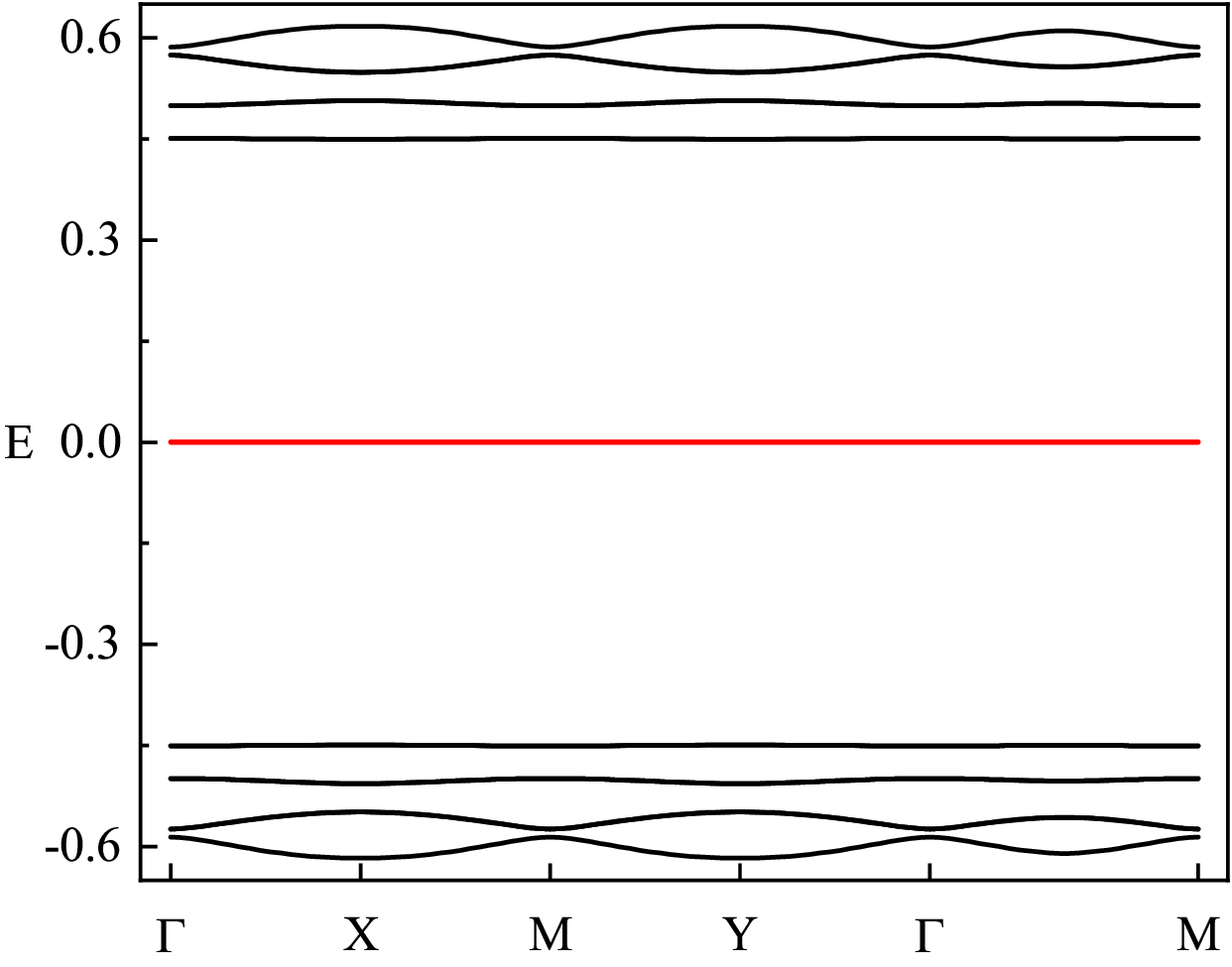, width=0.3\textwidth} \label{fig:Lattice_Bands2}
			}
			\subfigure[]{
				\epsfig{figure=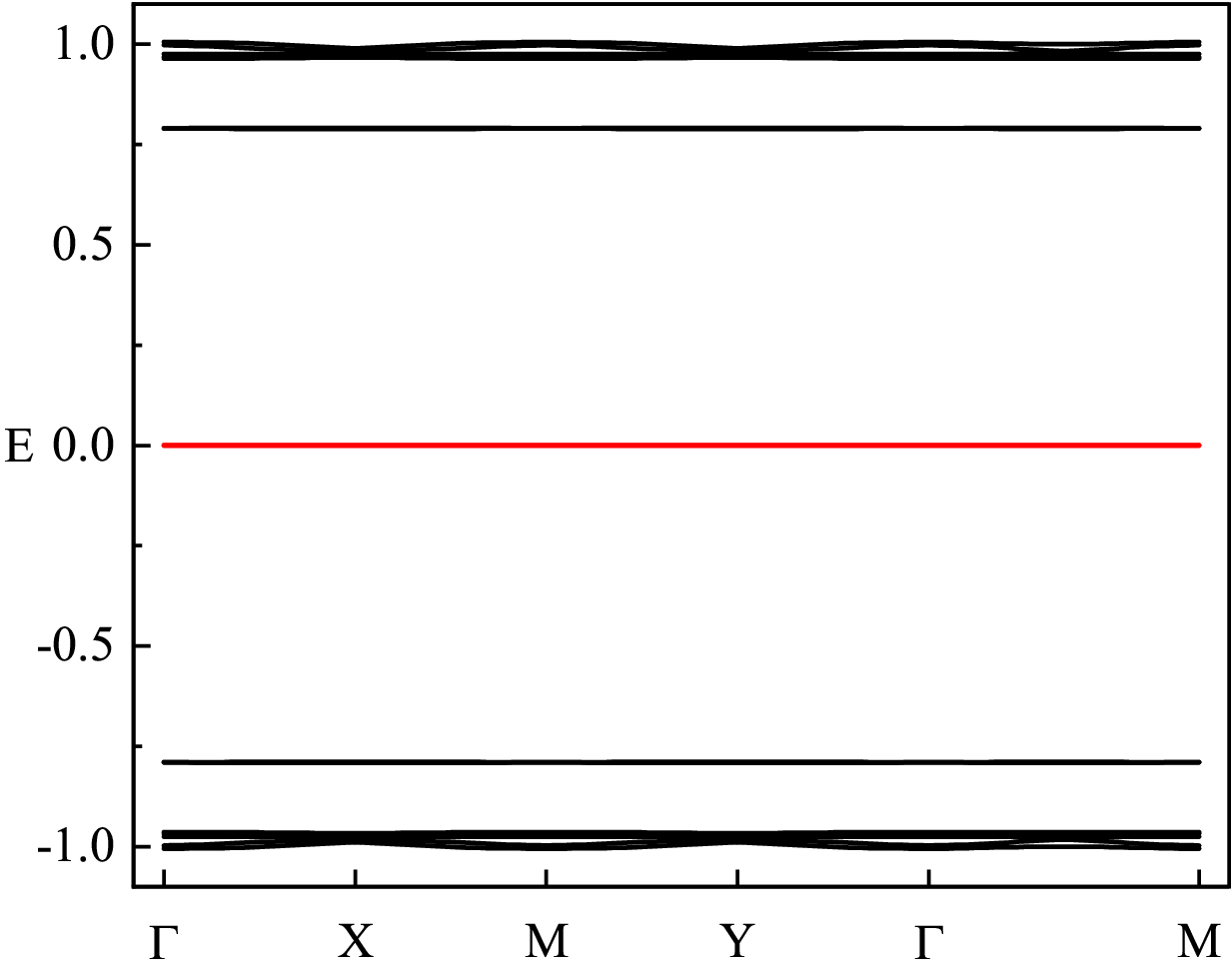, width=0.3\textwidth} \label{fig:Lattice_Bands3}
			}
		\end{minipage}
		
		\begin{minipage}[b]{0.95\textwidth}
			\centering
			\subfigure[]{
				\epsfig{figure=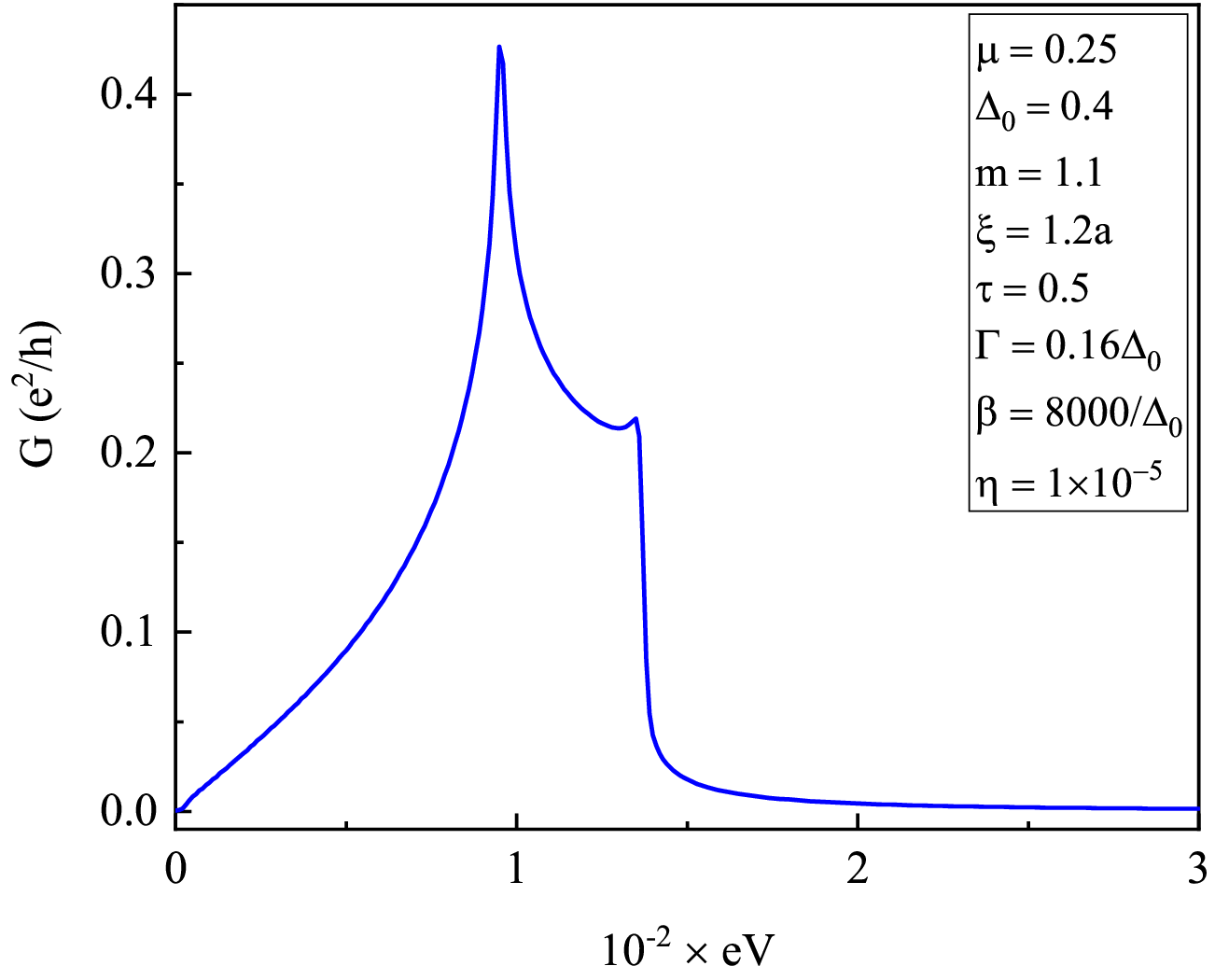, width=0.3\textwidth} \label{fig:LatticeG(V)1}
			}
			\subfigure[]{
				\epsfig{figure=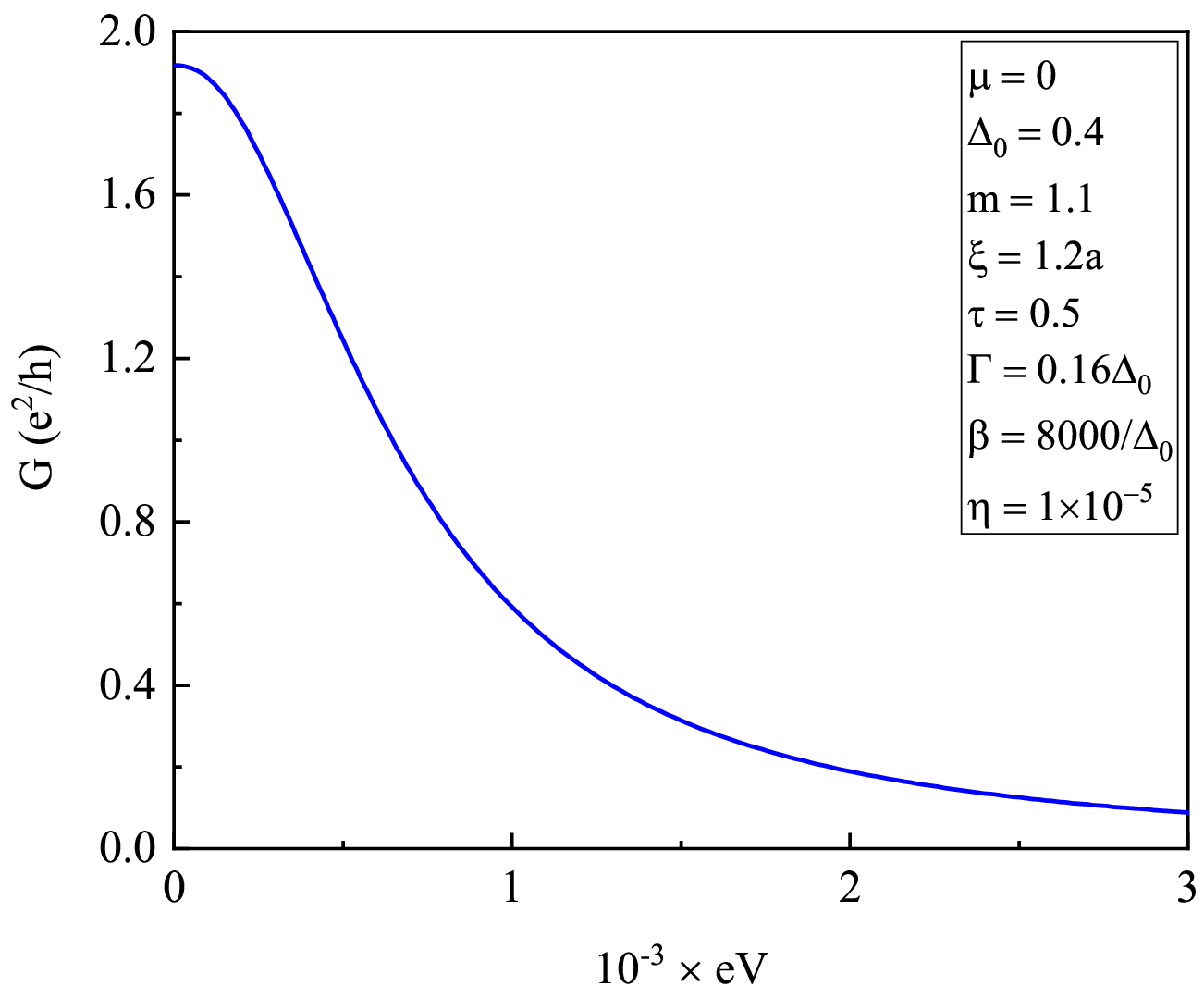, width=0.3\textwidth} \label{fig:LatticeG(V)2}
			}
			\subfigure[]{
				\epsfig{figure=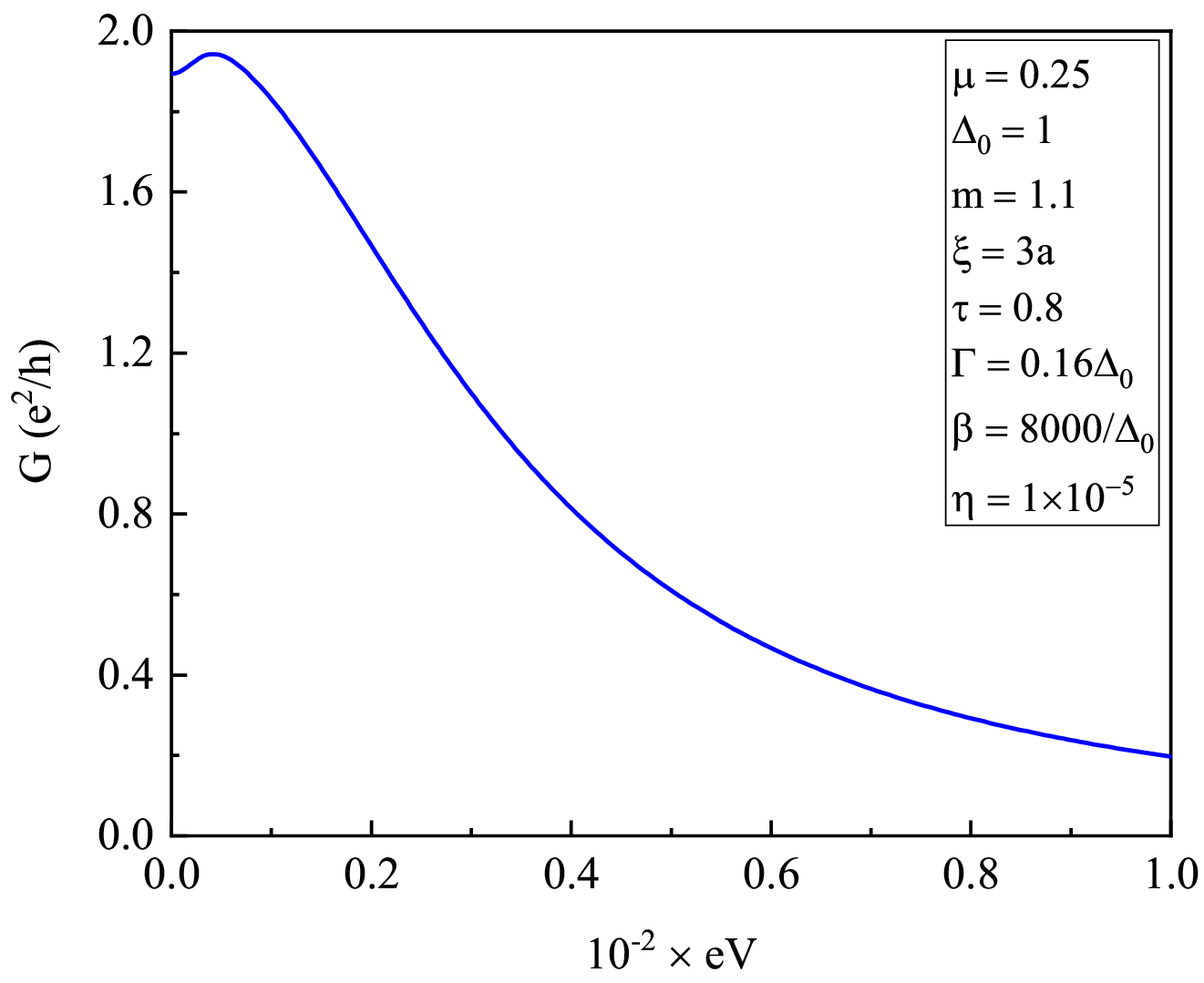, width=0.3\textwidth} \label{fig:LatticeG(V)3}
			}
		\end{minipage}
		
		\caption{(a)-(f) depict different cases of Majorana bands and their conductance at the vortex core center. The vortex lattice is constructed using a square lattice tight-binding model with a lattice constant of $a=1$, where the nearest neighbor hopping amplitude is set to $t=1$. A $30\times30$ magnetic unit cell is considered for all cases. The energy width is set to $\Gamma = 0.16\Delta_0$, the inverse temperature is $\beta = 8000/\Delta_0$, and the relaxation parameter is $\eta = 1\times10^{-5}$. In cases (b) and (c) the couplings between MZMs at different vortex cores are weak (in fact at $\mu=0$, an additional chiral symmetry occurs, causing the overlap amplitudes between distinct MZMs to vanish), and the corresponding ZBCP is nearly the quantized value. In contrast, for figure (a), the relatively strong couplings lead to a strong suppression of the conductance peak's height and a shift of the peak position to a finite energy.}
		\label{fig5}
	\end{figure*}
	However, as discussed above, we can gain insights of the MZM lattice from the Fano factor tomography introduced in Sec. \ref{sec:2}. In order to achieve a more accurate description of the vortex lattice, we use a periodic Hamiltonian for the Fu-Kane model in the presence of a vortex lattice. We take the vortex lattice approach proposed by Refs.\cite{PhysRevB.92.134519,PhysRevB.63.134509} and perform the band calculation on a square lattice. The band structure for a square vortex lattice and a generic chemical potential $\mu$ is shown in Fig. \ref{fig:Lattice_Bands1}. Besides the Majorana bands (red lines) around zero energy, there are bulk superconducting bands (black lines) with superconducting gaps. Tuning $\mu$ to the neutrality point ($\mu=0$) results in the vanishing of couplings between the MZMs, leading to a completely flat Majorana band, which is shown in Fig. \ref{fig:Lattice_Bands2}. This is because $H_{\text{FK}}$ exhibits an extra chiral symmetry generated by $\Pi=\tau_{z}\sigma_{z}$ at the neutrality point. As an important consequence, the overlap amplitudes $|t_{ij}|$ between distinct MZMs exactly vanish at this point\cite{PhysRevB.82.094504}. Fig. \ref{fig:Lattice_Bands3} shows that a larger pairing amplitude $\Delta_0$ results in a more localized MZM wave function and, consequently, smaller overlap amplitudes. Comparing Figure \ref{fig:LatticeG(V)1} with Figures \ref{fig:LatticeG(V)2} and \ref{fig:LatticeG(V)3}, we observe that the differential conductance of a MZM in the vortex lattice can be substantially reduced, even though the Majorana band exhibits a weak dispersion.

     \begin{figure}[!htbp]
		\begin{minipage}[b]{0.5\textwidth}
			\centering
			\subfigure[]{
				\epsfig{figure=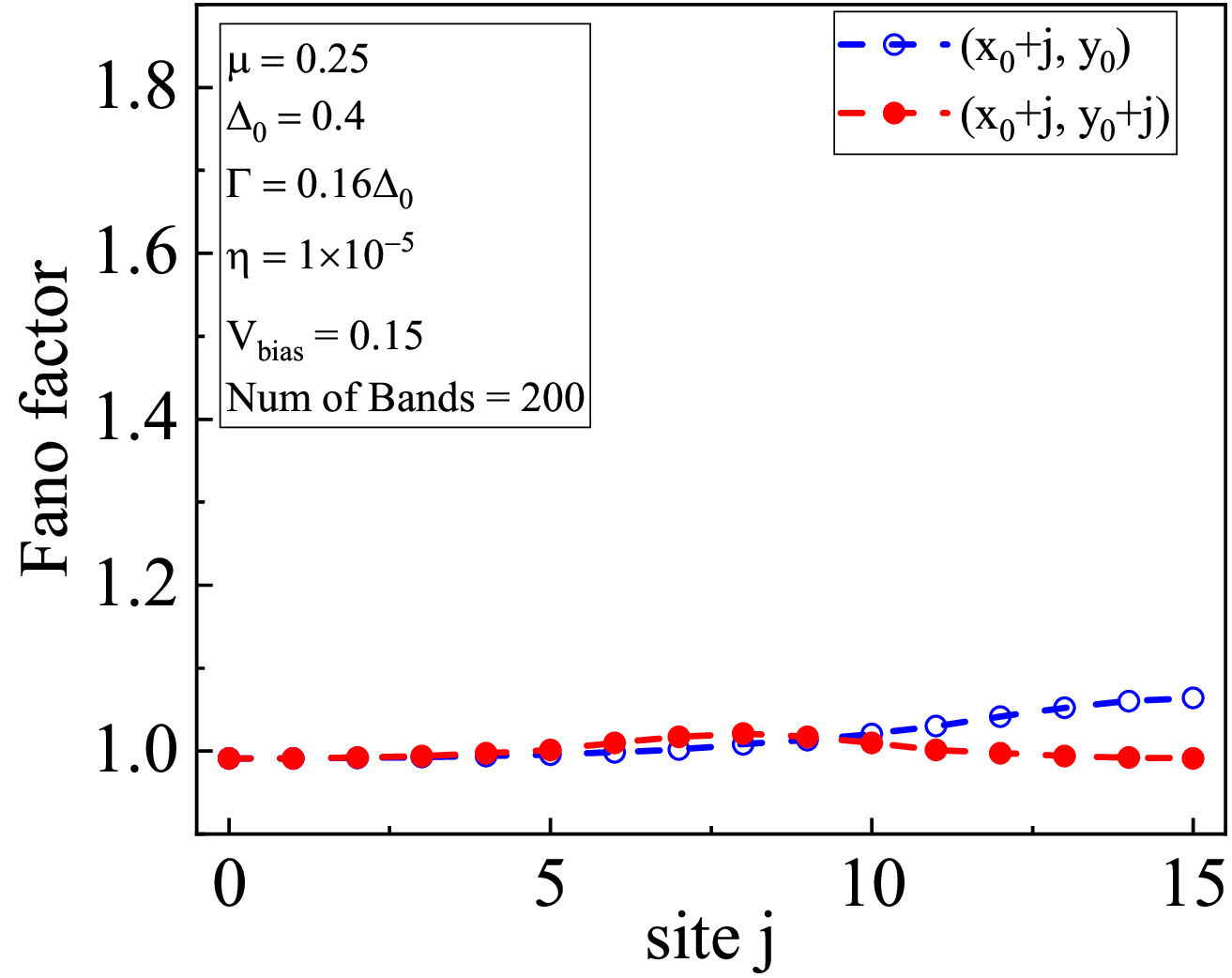, width=0.45\textwidth} \label{fig6:sub1}
			}
			\subfigure[]{
				\epsfig{figure=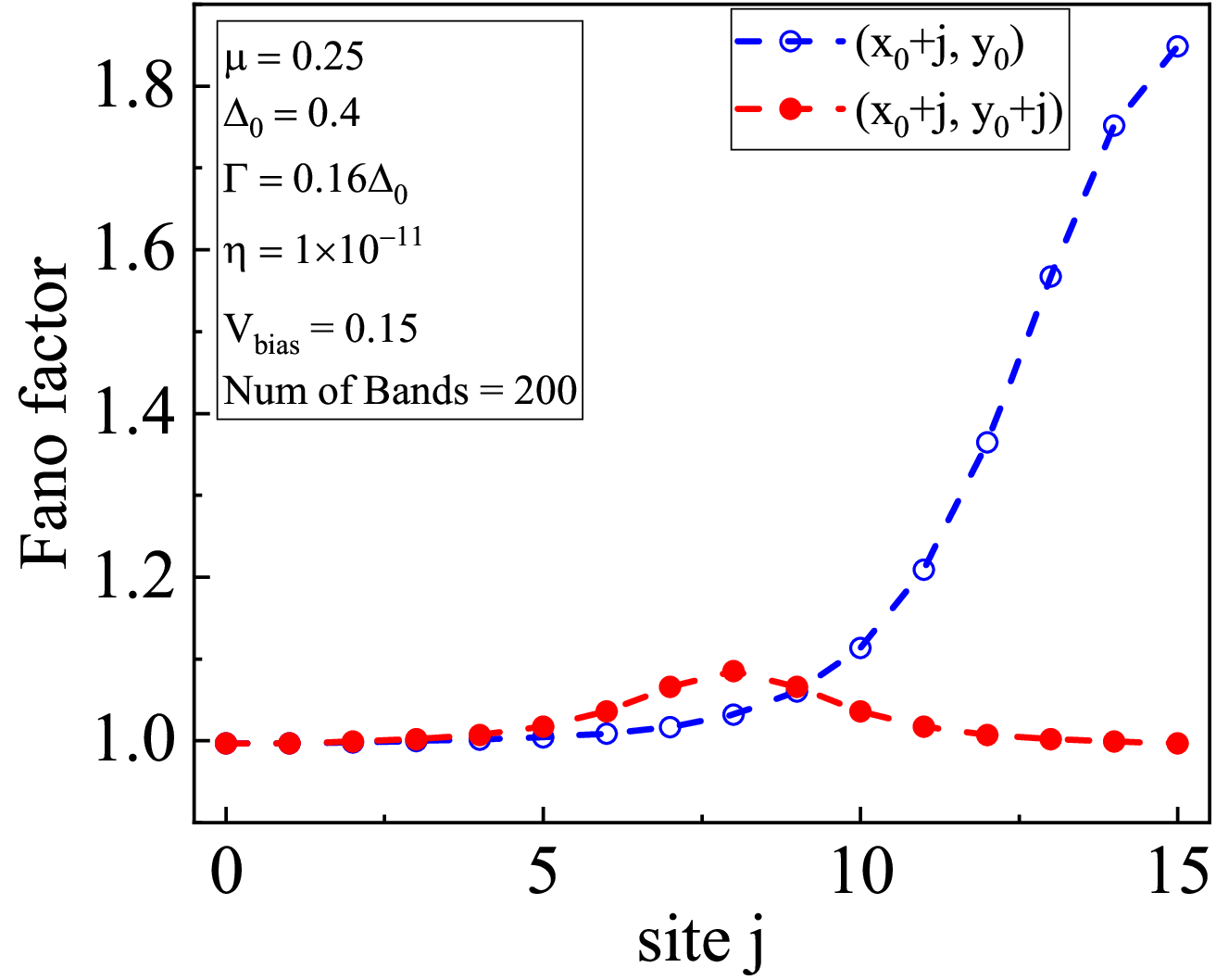, width=0.45\textwidth} \label{fig6:sub2}
			}
		\end{minipage}

  		\begin{minipage}[b]{0.5\textwidth}
			\centering
			\subfigure[]{
				\epsfig{figure=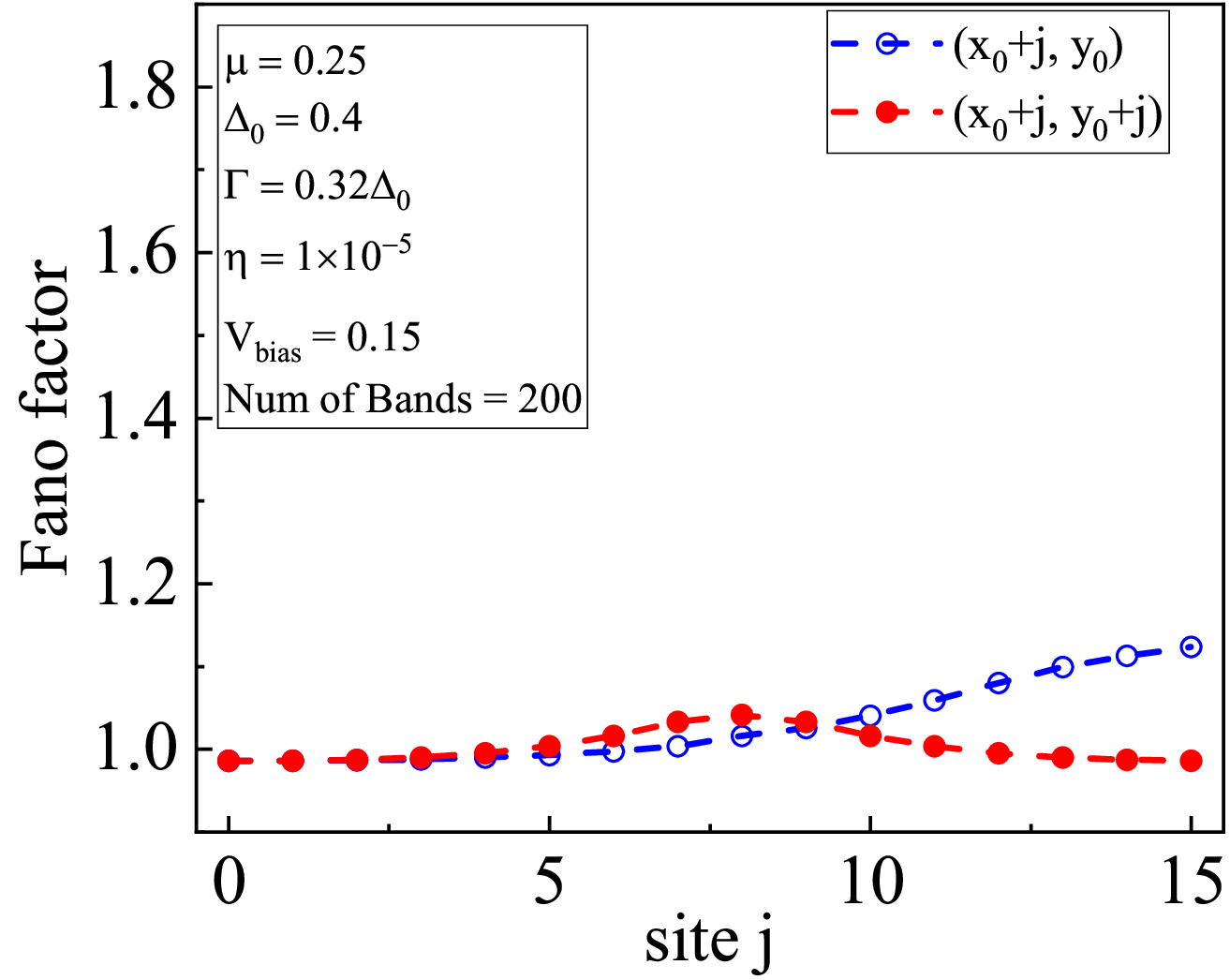, width=0.45\textwidth} \label{fig6:sub3}
			}
			\subfigure[]{
				\epsfig{figure=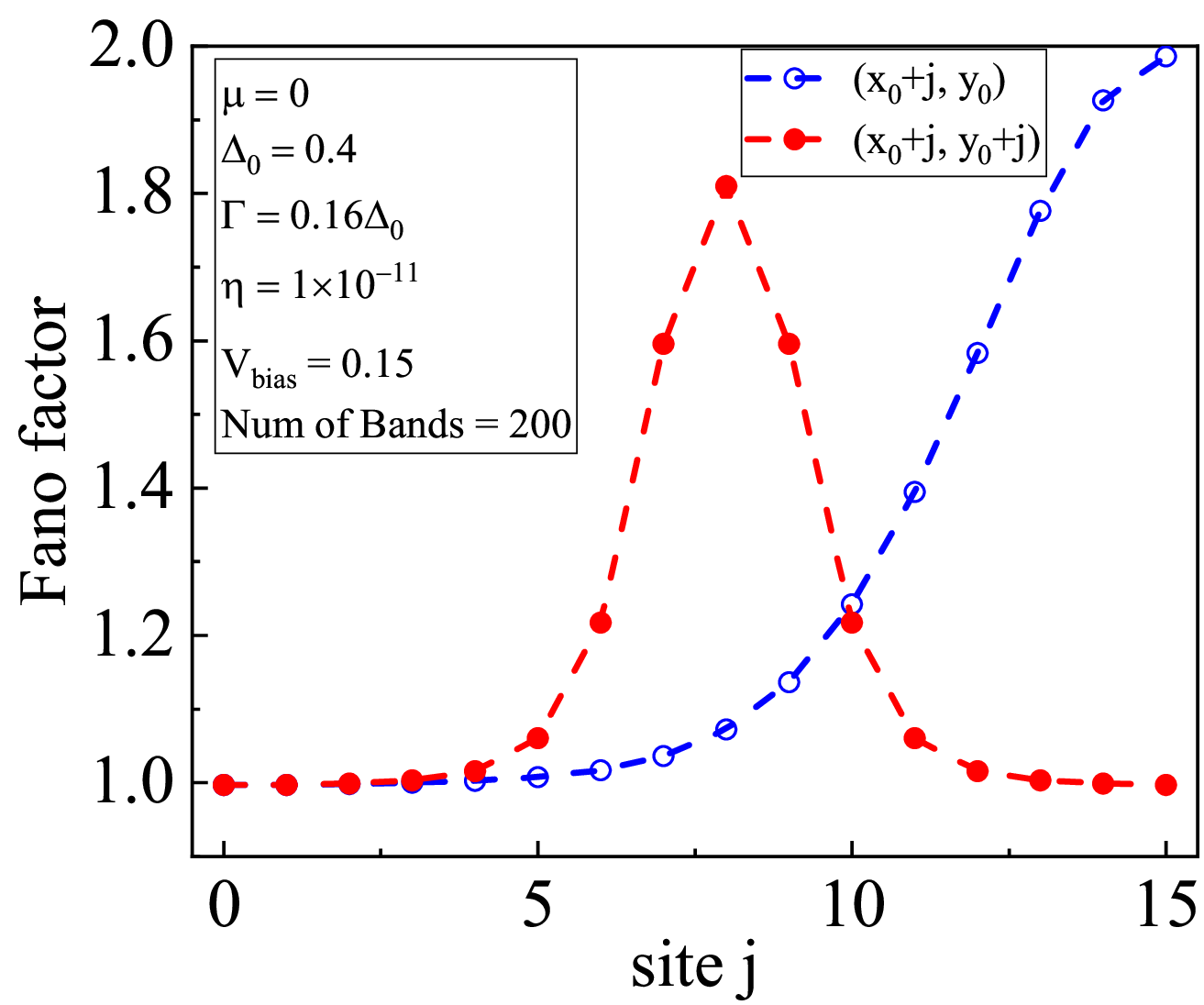, width=0.45\textwidth} \label{fig6:sub4}
			}
		\end{minipage}
		\caption{Fig. (a)-(d) show the spatially resolved Fano factors near vortex cores and their dependence on tunneling parameters. The coordinate $(x_0, y_0)$ represents the location of the vortex core center. In all cases, a pairing amplitude of $\Delta_0=0.4$ and a bias voltage $V_{\text{bias}}=0.15$ are used. The remaining parameters are specified as follows: (a) $\{\mu,\Gamma,\eta\}=\{0.25,0.16\Delta_0,1\times10^{-5}\}$, (b) $\{\mu,\Gamma,\eta\}=\{0.25,0.16\Delta_0,1\times10^{-11}\}$, (c) $\{\mu,\Gamma,\eta\}=\{0.25,0.32\Delta_0,1\times10^{-5}\}$, and (d) $\{\mu,\Gamma,\eta\}=\{0,0.16\Delta_0,1\times10^{-11}\}$. Figures (a), (b) and (c) are corresponding to the case shown in Fig. (\ref{fig:Lattice_Bands1}, \ref{fig:LatticeG(V)1}), while Figure (d) corresponds to Fig. (\ref{fig:Lattice_Bands2}, \ref{fig:LatticeG(V)2}). The values of the Fano factors plateau at $1$ in the vicinity of a vortex core, regardless of the tunneling parameter details. As the relaxation parameter $\eta$ approaches zero (cases (a) and (b)), and the energy width $\Gamma$ increases (cases (a) and (c)), the effect of the overlapping between MZMs becomes more apparent (the red lines), and Andreev reflections from the bulk states begin to contribute to the Fano factors (the blue lines), away from the vortex core.}\label{fig6}
    \end{figure}
	
	Furthermore, we calculate the spatially resolved Fano factors of the Fu-Kane model in the presence of the vortex lattice. As shown in Figures \ref{fig6:sub1}-\ref{fig6:sub4}, the Fano factors plateau at value of 1 near each vortex core, regardless of the tunneling parameter details (i.e., relaxation parameter $\eta$, tunneling energy width $\Gamma$ and chemical potential $\mu$ of the Fu-Kane model). Away from the vortices, the occurrence of Andreev reflection from the bulk states is enhanced by a larger tunneling energy width $\Gamma$ (as indicated by the blue lines in Fig. \ref{fig6:sub1} and \ref{fig6:sub3}). When the wave functions of the MZMs from different vortex cores overlap, the Fano factor is expected to be greater than 1 in accordance with Eq. (\ref{sec2:eqn3}). However, Eq. (\ref{sec2:eqn3}) is valid only when the relaxation parameter, $\eta$, approaches zero, where the single electron tunneling process is suppressed and only the Andreev reflection remains. Since the MZM wave function is small in that region, the behavior of the Fano factor is highly influenced by the finite relaxation parameter $\eta$. Additional cases of paired vortices and vortex lattices have been investigated in Appendix B. Spatially resolved Fano factors are less sensitive to the energy splitting of the Majorana bound states (or the dispersion of Majorana bands) compared to the differential conductance. Therefore, in cases where the overlap between the wave functions of the MZMs is not severe (but may still lead to a considerable energy splitting), Fano factor tomography may serve as a valuable tool to identify the existence of a well-separated MZM.
	
	\section{Discussion}
	The technique of tunneling measurement is commonly employed for the identification of Majorana zero modes (MZMs). A characteristic feature of an isolated MZM is the presence of a quantized zero-bias conductance peak. However, due to factors like finite temperature, as well as couplings with other low-lying bound states, the observation of this quantization becomes exceedingly challenging in realistic experiments. In a previous study\cite{PhysRevB.104.L121406}, the authors successfully demonstrated that current shot-noise spatial tomography with a metallic tip can distinguish Majorana bound state from other trivial zero-energy fermionic states in 1D nanowires. In the high voltage regime, the Fano factors reflect the local particle-hole asymmetry of the bound state. Since the wave function of the MZM is local particle-hole symmetric, the Fano factor $F(j)$ near it will plateau at 1. In this paper, we point out that this suppression of the Fano factor to half of the Poisson limit is a consequence of the existence of a perfect transparency peak in the Andreev reflection eigenvalue $T^{he}(E)$. The finite energy splitting $\varepsilon$ between these two MBSs primarily shifts the position of the transparency peak of $T^{he}(E)$ but has minimal effects on the Fano factor measured in the high voltage regime. When the wave funcions of these two MBSs begin to overlap, both of the energy width $\Gamma_0$ and $\bar{\Gamma}$ become non-zero, which generally reduces the height of the transparency peak and increases the Fano factor in the high voltage regime. As $\bar{\Gamma}$ approaches $\Gamma_{0}$ and the phase angle $\theta\rightarrow\pi/2$, the Andreev reflection eigenvalue becomes vanishingly small and the Fano factor recovers its classical value of 2. We also note that tunneling into one MBS resembles tunneling through a symmetric double barrier junction. In fact, in the high voltage regime the distribution of the transmission eigenvalues is the same universal bimodal function which leads to a suppression factor $1/2$ in both cases.
 
    In this paper, we apply the Fano factor tomography study to the 2D case. Our findings show that Fano factor tomography can effectively distinguish between MZM and CdGM bound states in the single vortex case. In the vicinity of a MZM, a flat plateau of $F=1$ is observed, in stark contrast to the case of trivial in-gap states (such as CdGM bound states and YSR bound states), where strong spatial oscillations of $F$ between values of 1 and 2 are obtained. In periodic vortex lattices, the MZMs within each vortex interact with each other, giving rise to a low-lying in-gap band. Our study demonstrates that, despite of the weak dispersion of the Majorana band, the differential conductance of an individual MZM can be significantly reduced. Conversely, the spatially resolved Fano factors exhibit lesser sensitivity to the energy splitting of the Majorana bound states and consistently plateau at 1 near each vortex core. As a result, Fano factor tomography can serve as an additional tool for identifying the presence of well-separated MZMs.
	
	We further examine the influence of the single-electron tunneling processes on Fano factor tomography measurements. Single-electron tunneling can occur due to relaxation processes of in-gap bound states into the BCS continuum, which can be characterized by a finite relaxation parameter, $\eta$. It was found that the Fano factor can indicate local particle-hole asymmetry of bound states only when $\eta\ll\Gamma_j\ll eV$. An essential condition for this is a distinct separation between the density of states of in-gap states and bulk states. In experimental settings, temperature reduction can suppress relaxation from bound states to the quasiparticle continuum\cite{PhysRevLett.115.087001}, thereby reducing $\eta$.
	
	Recently, an ordered and tunable lattice of MZMs has been discovered in the iron pnictide compound $\ce{LiFeAs}$\cite{doi:10.1038/s41586-022-04744-8}. Our results can facilitate the identification of isolated MZMs within the lattice and foster further experimental advancements of STM shot-noise experiments in the field of Majorana fermions.
	
	\section*{Acknowledgments}
This work is supported by the Ministry of Science and Technology  (Grant No. 2022YFA1403900), the National Natural Science Foundation of China (Grant No. NSFC-11888101, No. NSFC-12174428, No. NSFC-11920101005), the Strategic Priority Research Program of the Chinese Academy of Sciences (Grant No. XDB28000000, XDB33000000), the New Cornerstone Investigator Program, and the Chinese Academy of Sciences through the Youth Innovation Promotion Association (Grant No. 2022YSBR-048).

	\section*{Appendix A: CdGM bound states}
	For an ordinary s-wave superconductor, one can just replace $\hat{h}$ in Eq. (\ref{sec2:eqn4}) with $\hat{h}=\frac{\bm{p}^2}{2m}-\mu$. Here we use a simplest square lattice model to simulate a 2D SC with a single vortex with the dispersion relation as $\hat{h}=-2t(\cos{k_x}+\cos{k_y})+4t-\mu$.
	\begin{figure}[!htbp]
		\centering
		\subfigure[]{
			\epsfig{figure=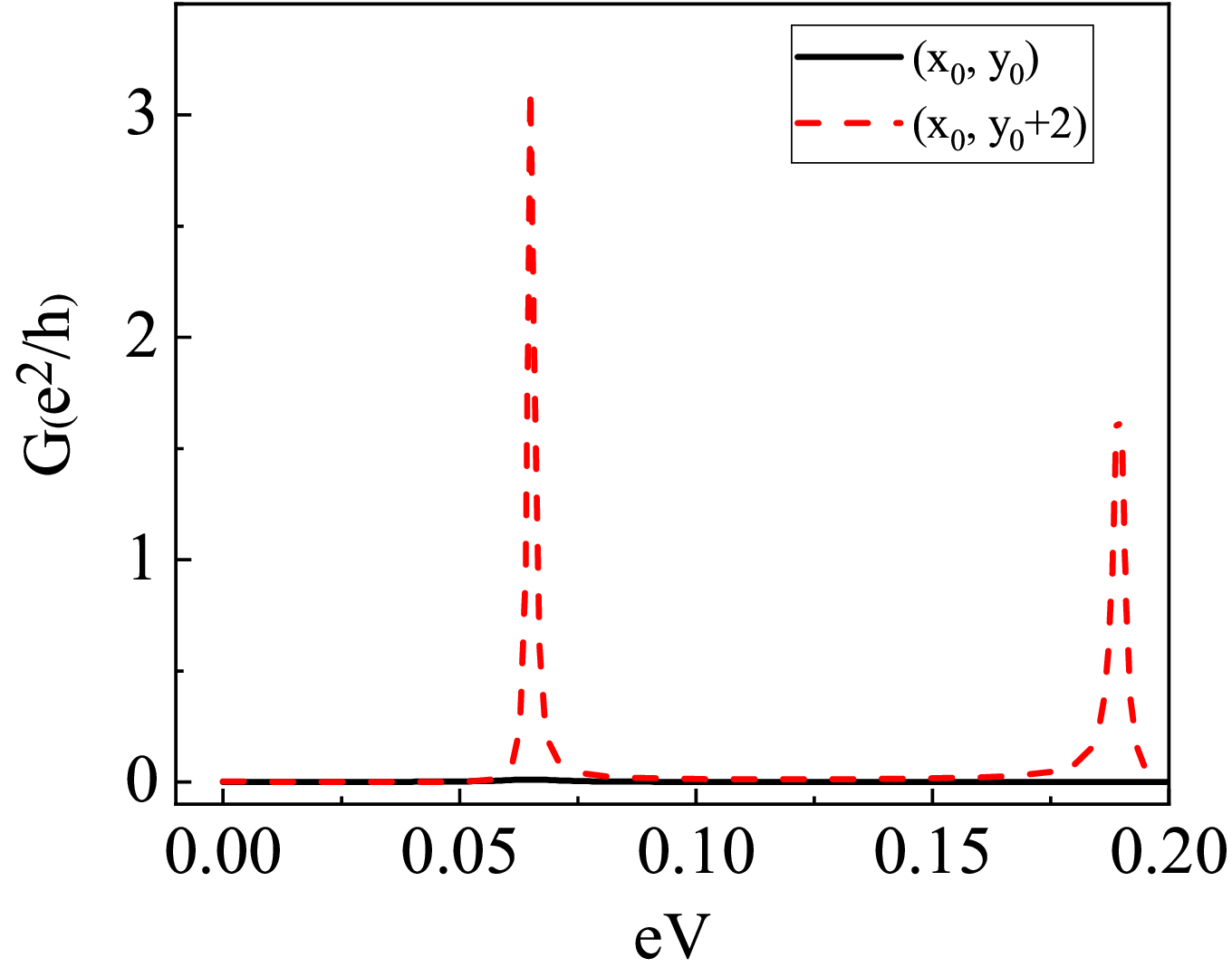, width=0.23\textwidth} \label{figappendixA1:sub1}
		}
  		\subfigure[]{
			\includegraphics[width=0.22\textwidth]{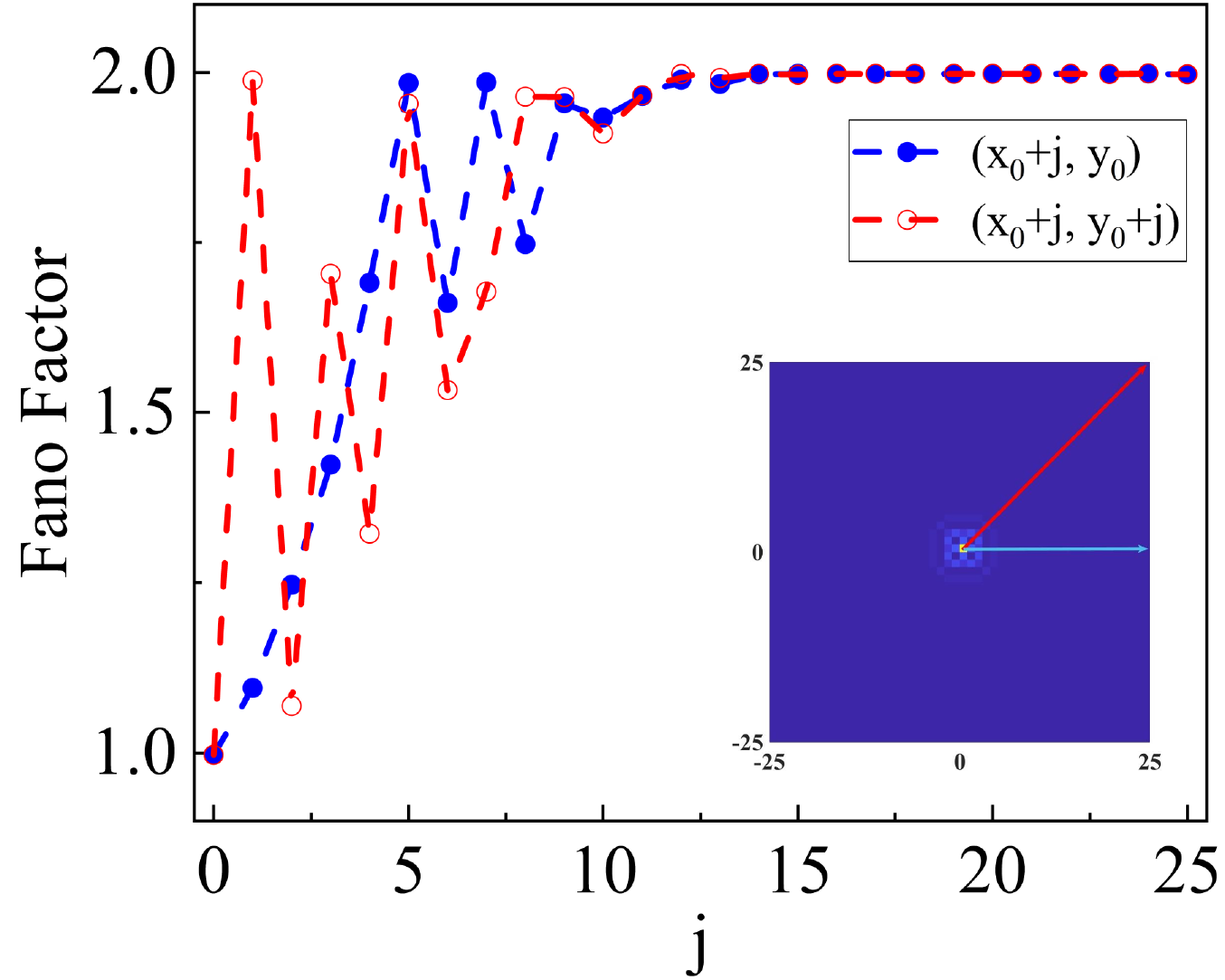} \label{figappendixA1:sub2}
		}
		\caption{Differential conductance and spatially resolved Fano factors are shown for a single vortex with CdGM bound states. The parameters in the simulation are $\{t,\,\mu,\,\Delta_0,\,\beta,\,\Gamma,\,\eta\}=\{1.5,\,3.5,\,0.8,\,1\text{e}4,\,0.128,\,1\text{e-5}\}$. The inset in panel (b) shows the lowest lying eigenstate of the CdGM bound states, with arrows indicating the scanning directions. The vortex core is located at coordinates $(x_0, y_0)$, and Fano factors were measured by setting a voltage bias of $V_{\text{bias}}$ = 0.15 between the energy of the lowest bound state and the second lowest state.}\label{figappendixA1}
	\end{figure}
    The differential conductance and Fano factor tomography are presented in Fig. \ref{figappendixA1}. At first sight of Fig. \ref{figappendixA1:sub1}, one may feel strange about the vanishingly small conductance at the vortex core center. This is because we are working in the strong tunneling regime where the naive expectation that the differential conductance is proportional to the local density of states is no longer valid. In this case, it is the Andreev reflection that dominates in the tunneling current. For a SC with full spin rotation symmetry, the differential conductance near its in-gap bound states resonance can be given by,
    \begin{equation}\label{appendixAEq:1}
    G(\omega)=\frac{2e^{2}}{h}\frac{8u^{2}v^{2}\Gamma^{2}\varepsilon^{2}}{\left[\omega^{2}-\left(\varepsilon^{2}-(u^{2}+v^{2})^{2}\Gamma^{2}/4\right)\right]^{2}+(u^{2}+v^{2})^{2}\Gamma^{2}\varepsilon^{2}},
    \end{equation}
    where $\varepsilon$ is the energy of the lowest CdGM bound states. At the vortex core center, one of the $u,\,v$ components of the CdGM bound states is zero\cite{CAROLI1964307}, leading to a vanishingly small conductance.

    Even though in this case the number of the lowest bound states are doubled due to the $\text{SU}(2)$ degeneracy, the Fano factor tomography still has a strong spatial oscillation around the vortex core which is completely different with the MZM case.
    
	\section*{Appendix B: Fano factors of a pair of vortices and vortex lattices}
	For a single-orbital Hamiltonian, a zero-energy bound state can be described by the Green's function, given by
	\begin{equation}\label{appendixBEq:1}
		g_{S;j,j}^{R}(\omega)=\frac{\phi_{+}(j)\phi_{+}^{\dagger}(j)+\phi_{-}(j)\phi_{-}^{\dagger}(j)}{\omega+i\eta}.
	\end{equation}
	Here,  $\phi_{+}=\left[u_{\uparrow}(j),\,u_{\downarrow}(j),\,v_{\downarrow}(j),\,-v_{\uparrow}(j)\right]^{T}$ represents the wave function associated with the zero-energy bound state, while $\phi_{-}=\tau_{y}\sigma_{y}\mathcal{K}\phi_{+}(j)$ is its particle-hole symmetric counterpart. Our primary focus is on the high-voltage regime, where $eV\gg\Gamma_j$, allowing us to consider $V\rightarrow\infty$. Additionally, we assume a temperature of $T=0$. When the relaxation parameter $\eta\rightarrow 0^+$ and the Nambu-spinor $\phi_{+}$ can be gauged into a real one, the integral of the current and the noise can be rigorously computed\cite{PhysRevB.104.L121406}. Consequently, the Fano factor is exact and given by Eq. (\ref{sec2:eqn3}). Remarkably, Eq. (\ref{sec2:eqn3}) remains accurate even for general complex Nambu-spinors, in agreement with previous findings\cite{PhysRevB.104.L121406}.
	
	Finite $\eta$, which physically could arise from the relaxation of the in-gap bound state to the BCS quasiparticle continuum, allows for single electron tunneling processes. In contrast to Andreev reflection, which transports Cooper pairs and has an effective charge of 2e, single-electron tunneling has an effective transport charge of 1e. Since the Fano factor reflects the effective transport charge (at least in the Poisson limit), single-electron tunneling can generally result in its reduction. Using Eq. (\ref{appendixBEq:1}) as input, we can follow the calculation in Ref.\cite{PhysRevB.104.L121406} to obtain the Fano factor in the limit of $V\rightarrow\infty$. The full expression is cumbersome, so we will only consider two limits here. When $\eta\ll\Gamma_j$, Andreev reflection dominates, resulting in the Fano factor being
	\begin{equation}\label{appendixBEq:2}
		F\simeq1+\left(\frac{\sum_{\sigma}u_{\sigma}^{2}-v_{\sigma}^{2}}{\sum_{\sigma}u_{\sigma}^{2}+v_{\sigma}^{2}}\right)^{2}\left(1-A\frac{\eta}{\Gamma_{j}}\right)+O\left(\frac{\eta}{\Gamma_{j}}\right)^{2},
	\end{equation}
	where the factor $A$ is always positive and is given by
	\begin{equation}\label{appendixBEq:3}
		A=4+\frac{\left(\sum_{\sigma}u_{\sigma}^{2}+v_{\sigma}^{2}\right)^{2}}{2\left(\sum_{\sigma}u_{\sigma}^{2}\right)\left(\sum_{\sigma}v_{\sigma}^{2}\right)}.
	\end{equation}
	As expected, a finite relaxation parameter $\eta$ would lead to a reduction of the Fano factor from Eq. (\ref{sec2:eqn3}). In the reverse limit of $\eta\gg\Gamma_j$, the Fano factor is
	\begin{equation}\label{appendixBEq:4}
		F\simeq1-\left(\frac{\sum_{\sigma}u_{\sigma}^{2}-v_{\sigma}^{2}}{\sum_{\sigma}u_{\sigma}^{2}+v_{\sigma}^{2}}\right)^{2}\frac{\Gamma_{j}}{\eta}+O\left(\frac{\Gamma_{j}}{\eta}\right)^{2}.
	\end{equation}
	In this case, single-electron tunneling dominates, and as $\eta\gg\Gamma_j$, the transmission transparency is low, resulting in $F\simeq1$, which is the result of single-electron transport in the Poisson limit. As we can see, the Fano factor become featureless in this case. To ensure effective Fano factor tomography, we must operate in the regime where $\eta\ll\Gamma_j\ll eV$.
	
\begin{figure}[!htbp]
		\centering
		\subfigure[]{
			\epsfig{figure=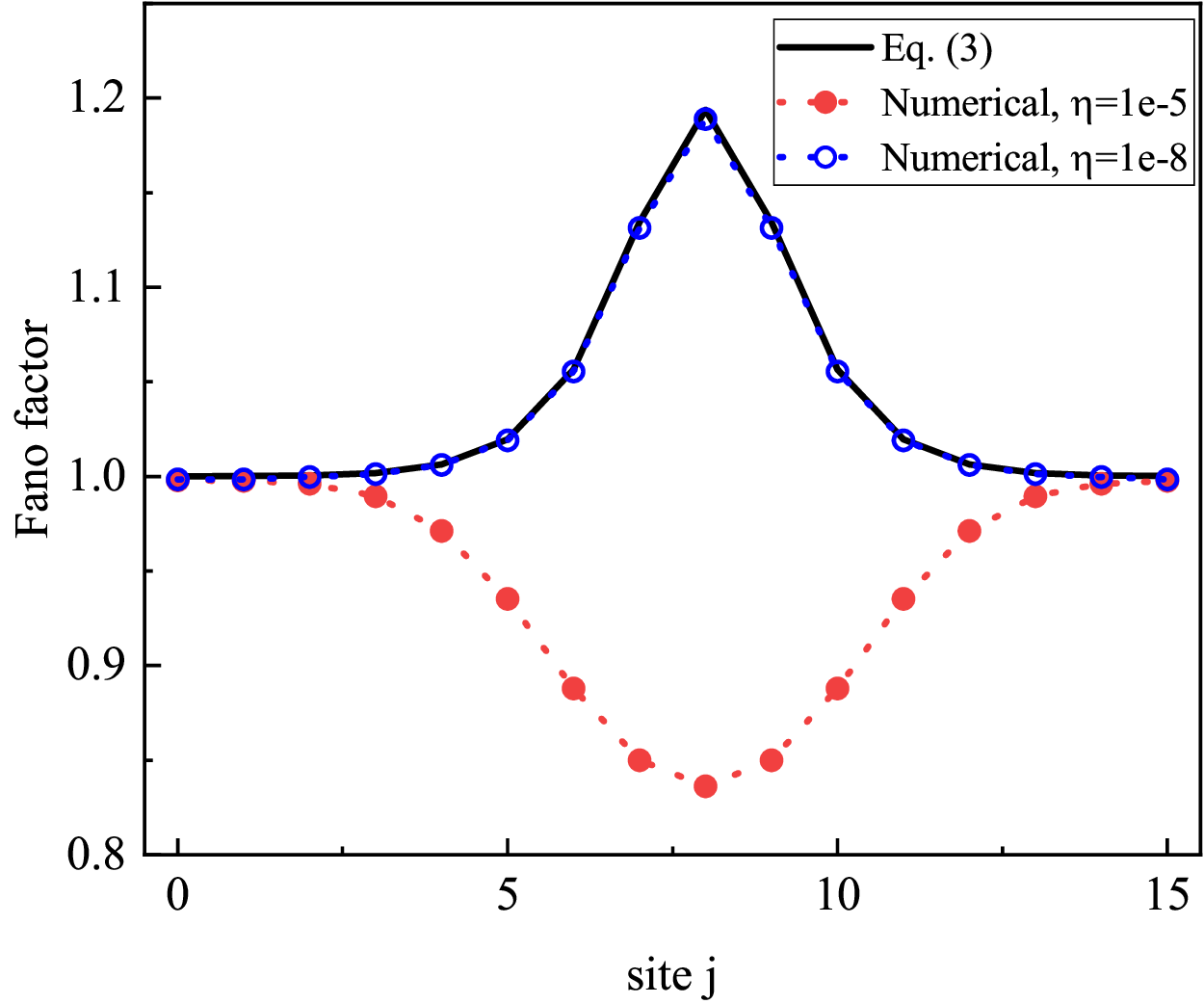, width=0.23\textwidth} \label{fig:F(j)TwoVortices_mu0.25}
		}
		\subfigure[]{
			\epsfig{figure=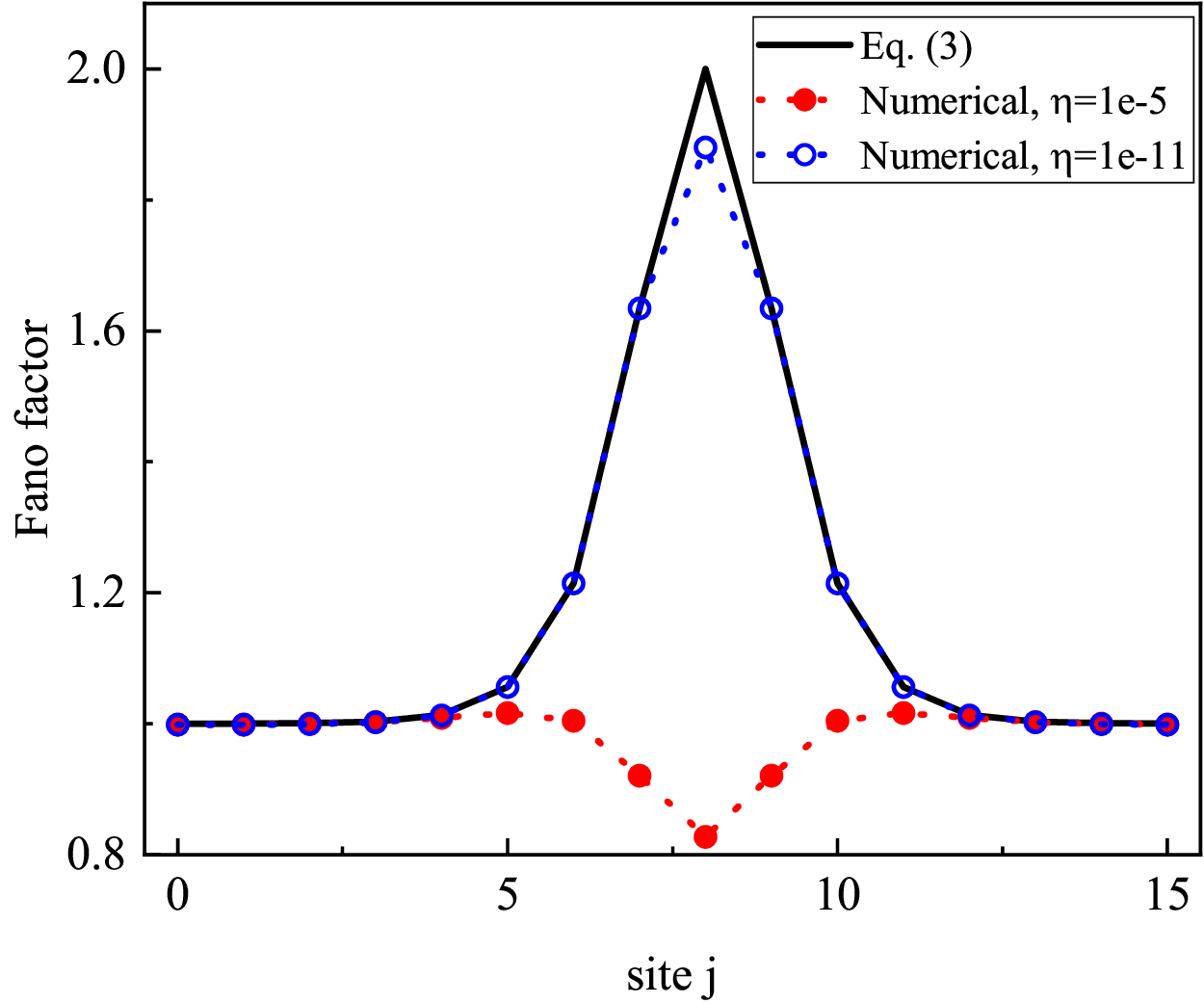, width=0.23\textwidth} \label{fig:F(j)TwoVortices_mu0}
		}
		\caption{This figure illustrates the spatially resolved Fano factors along the connected line of the two vortices in a Fu-Kane model. We use a square lattice to perform the numerical simulation, and the contribution is solely considered from the two MBSs. The figure displays the Fano factors for two different values of the chemical potential: (a) $\mu=0.25$, and (b) $\mu=0$. The solid lines represent the analytical results obtained from Eq. (\ref{sec2:eqn3}), while the data points correspond to the results of numerical simulations performed with varying relaxation parameter values $\eta$.}
		\label{figappendixB1}
	\end{figure}

	According to Eq. (\ref{appendixBEq:2}), when the amplitudes of the wave function of bound states at site $j$ are small (correspondingly, $\Gamma_j$ would be small), the finite relaxation parameter will exert a significant influence on the Fano factor. Figure \ref{figappendixB1} illustrates the measured Fano factors of a Fu-Kane model with two vortices, obtained as the STM tip sweeps from one vortex to the other. To facilitate a direct comparison with Eq. (\ref{sec2:eqn3}), we neglected the contribution from the bulk states in the simulation. It can be observed from Figure \ref{figappendixB1} that when the tip is positioned near the mid-point between these two vortices, where the Majorana wave function is small, a significant deviation from the prediction of Eq. (\ref{sec2:eqn3}) arises due to the finite relaxation parameter, $\eta$. The numerical simulation coincides with the analytical prediction only when $\eta$ is extremely small.
	
\begin{figure}[!htbp]
		\begin{minipage}[b]{0.5\textwidth}
			\centering
			\subfigure[]{
				\epsfig{figure=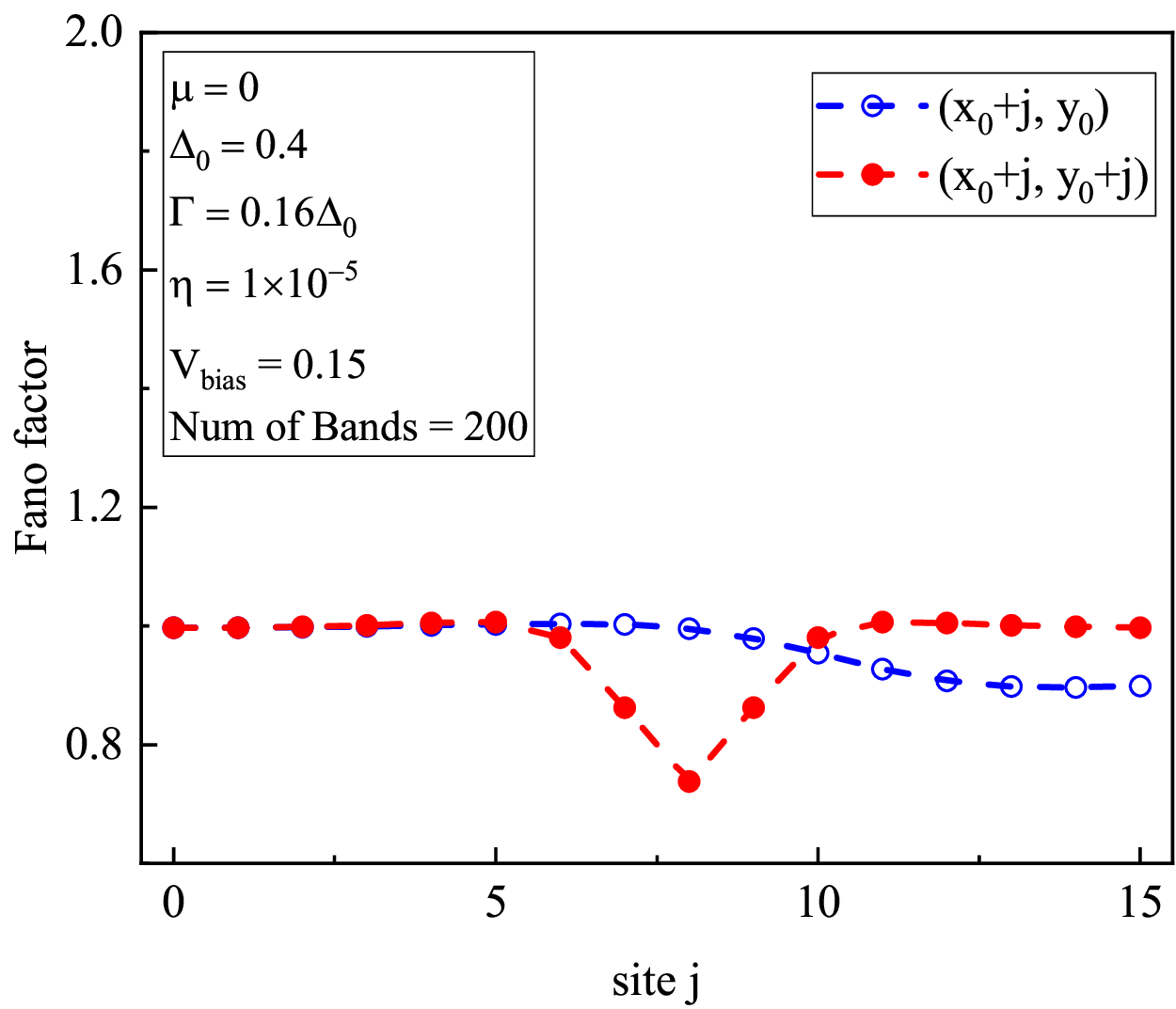, width=0.45\textwidth} \label{fig:F(j)2_200Bands,1e-5}
			}
			\subfigure[]{
				\epsfig{figure=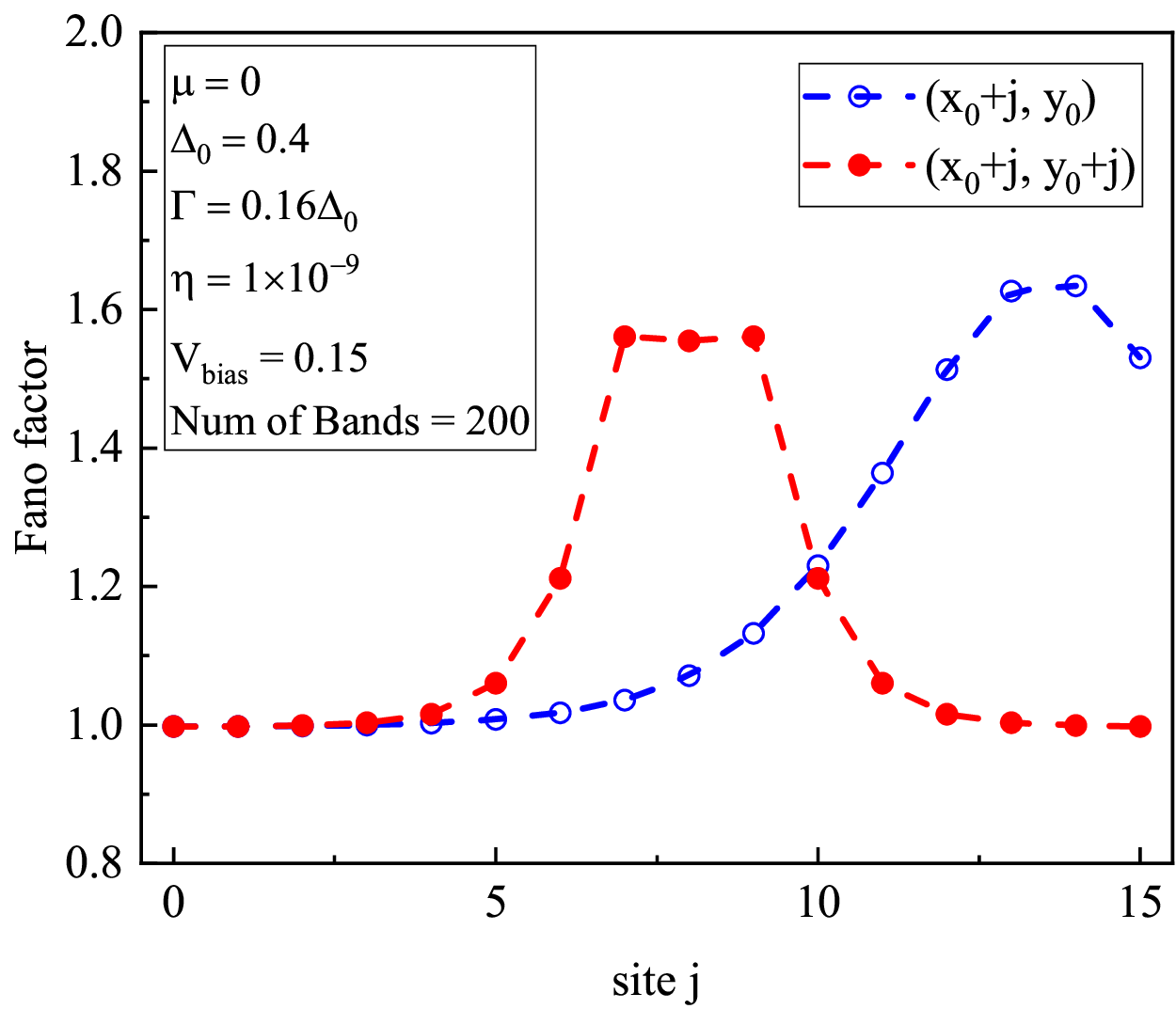, width=0.45\textwidth} \label{fig:F(j)_2_eta_10^-9}
			}
		\end{minipage}
		
		\begin{minipage}[b]{0.5\textwidth}
			\centering
			\subfigure[]{
				\epsfig{figure=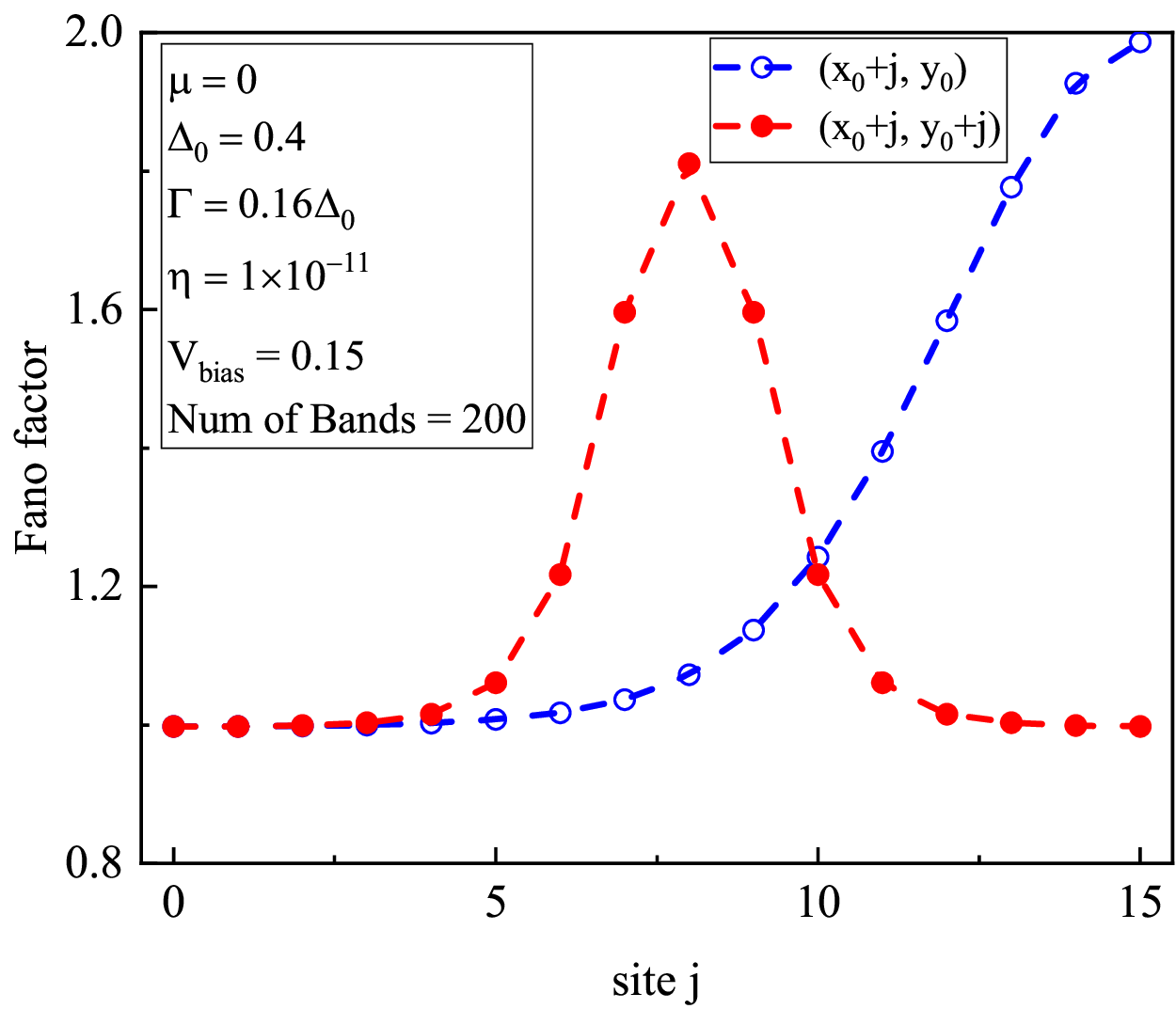, width=0.45\textwidth} \label{fig:F(j)_2_eta_10^-11}
			}
			\subfigure[]{
				\epsfig{figure=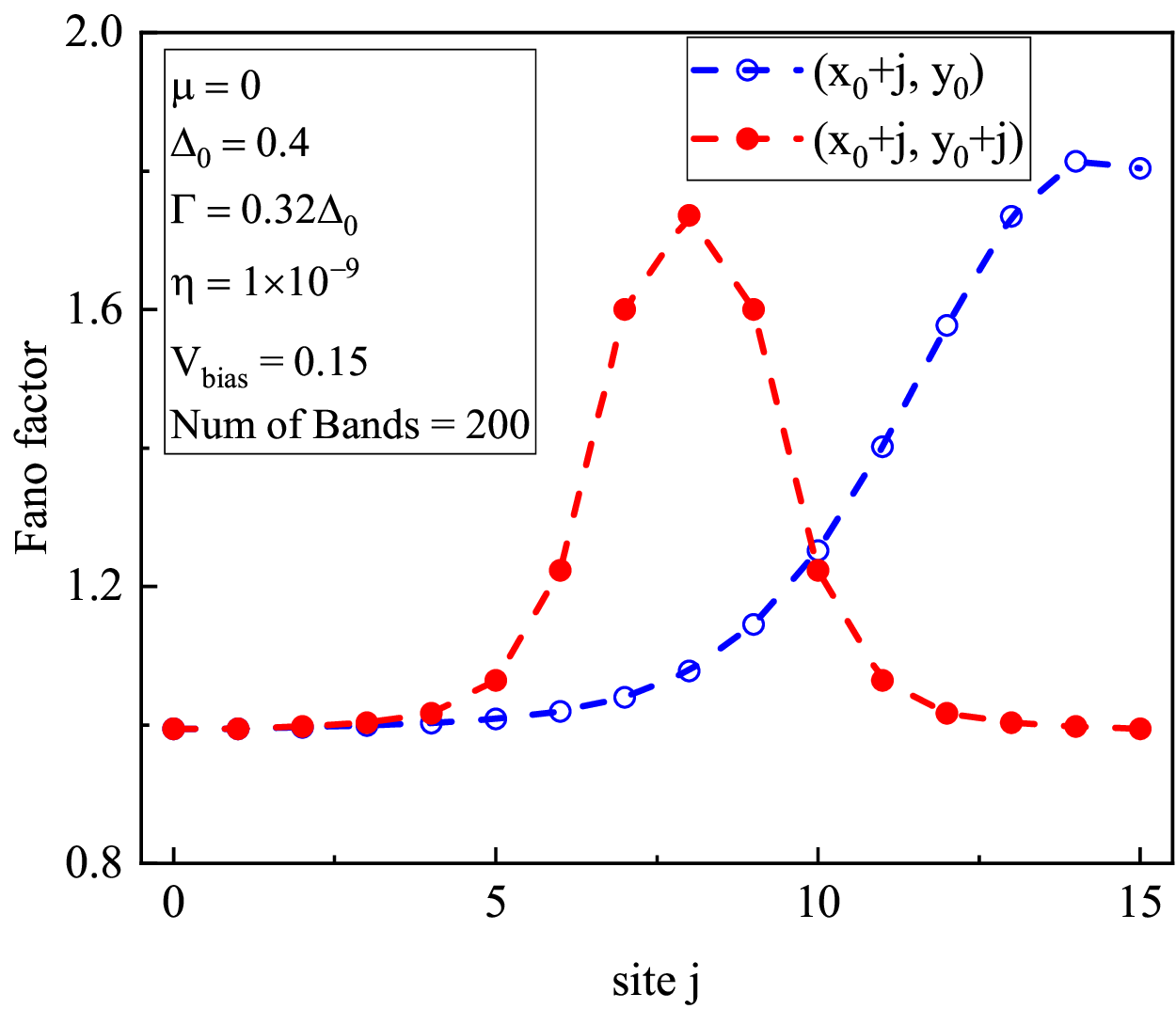, width=0.45\textwidth} \label{fig:LagerGamma_F(j)_2_eta_10^-9}
			}
		\end{minipage}
		\caption{These figures illustrate the spatially resolved Fano factors of a Fu-Kane model in the presence of a vortex lattice, with a fixed chemical potential of $\mu=0$ under different tunneling parameters. The coordinate $(x_0, y_0)$ represents the location of the vortex core center.}
		\label{figappendixB2}
	\end{figure}

	We calculated the Fano factor of the MZM vortex lattice under other parameters. As demonstrated in Fig. \ref{figappendixB2}, the behavior of the spatially resolved Fano factors is similar with the results shown in the maintext. Near each vortex core, the Fano factors plateau at 1 regardless of the tunneling details. Away from the vortex core the Fano factor is sensitive to the relaxation parameter $\eta$ and the energy width $\Gamma$ as shown in Fig. \ref{fig5} and Fig. \ref{figappendixB2}. As the relaxation parameter $\eta$ approaches zero, and the energy width $\Gamma$ increases, the effect of
    the overlapping between MZMs starts to become apparent.
	
	\section*{Appendix C: Majorana toy model and scattering method}
	Here we provide a scattering scheme to understand the behavior of the Fano factors. We examine a pair of Majorana bound states (MBSs) using a single spinless electrode measurement setup. The Majorana Hamiltonian is given by $H_M=i\varepsilon \gamma_1\gamma_2$. The unitary scattering matrix $S(E)$ can be written as
	\begin{equation}\label{appendixCEq:1}
		S(E)\equiv\begin{pmatrix}s^{ee} & s^{eh}\\
			s^{he} & s^{hh}
		\end{pmatrix}=1+2\pi iW^{\dagger}\left(H_{M}-E-i\pi WW^{\dagger}\right)^{-1}W,
	\end{equation}
	with $W$ the matrix that describes the coupling of the scatterer (Hamiltonian $H_M$) to the tip. In this toy model, we have
	\begin{equation}
		W=\begin{pmatrix}w_{0} & w_{0}\\
			\bar{w}e^{i\theta} & \bar{w}e^{-i\theta}
		\end{pmatrix},\qquad H_{M}=\begin{pmatrix}0 & i\varepsilon\\
			-i\varepsilon & 0
		\end{pmatrix}.
	\end{equation}
	The expression for $H_M$ is in the basis $\left\{ \Phi_{1},\Phi_{2}\right\} $ of the two MBSs, while $W$ is the coupling matrix in the basis $\left\{ \Phi_{e}^{tip},\Phi_{h}^{tip}\right\} $ of propagating electron and hole modes in the tip. We assume that the tip may couple to bound states 1 and 2 simultaneously due to the overlap of the two MBSs in real space. Without loss of generality, we can make $w_0$ purely real by adjusting the phase of the basis state in the tip.
	
	The general expressions for the time averaged current $\bar{I}$ and shot noise $S$ in the zero-temperature limit, in terms of the scattering matrix elements,  are\cite{PhysRevLett.101.120403,PhysRevB.53.16390}
	\begin{align}
		\bar{I} & =\frac{e}{h}\int_{0}^{eV}\mathrm{d}E\left(1-T^{ee}(E)+T^{he}(E)\right)\\
		S & =\frac{e^2}{h}\int_{0}^{eV}\mathrm{d}E\mathcal{P}(E),
	\end{align}
	with the definitions
	\begin{align}
		\mathcal{P}(E) & =T^{ee}(1-T^{ee})+T^{he}(1-T^{he})+2T^{ee}T^{he}\\
		T^{\alpha\beta}(E) & =\left|s^{\alpha\beta}(E)\right|^{2},\qquad\alpha,\beta\in\left\{ e,h\right\}.
	\end{align}
	In this simple case, the unitary condition tells us that $T^{ee}+T^{he}\equiv1$, and we can rewrite the current and the shot-noise in a more compact form,
	\begin{align}
		\bar{I} & =\frac{2e}{h}\int_{0}^{eV}\mathrm{d}E\,T^{he}(E)\label{appendixCEq:7}\\
		S & =\frac{8e^{2}}{h}\int_{0}^{eV}\mathrm{d}E\,T^{he}(E)\left(1-T^{he}(E)\right).\label{appendixCEq:8}
	\end{align}
	Here $T^{he}(E)$ is exactly the Andreev reflection eigenvalue with  $T^{he}(E)\in[0,1]$. For a trivial NS junction, in the zero-temperature, zero voltage limit, the shot-noise is given by\cite{PhysRevB.49.16070}
	\begin{equation}
		S_{NS}=8e|V|\frac{e^{2}}{h}\sum_{n}T_{n}^{he}\left(1-T_{n}^{he}\right),
	\end{equation}
	and the current $I=G_{NS}V=\frac{2e^2V}{h}\sum_nT_{n}^{he}$. These Andreev reflection eigenvalues are all evaluated at $E=0$. Usually for a trivial NS junction, all these eigenvalues are small (i.e., in the so-called Poisson limit of transport), and the Fano factor $F=2$ is exactly the effective charge of the current carriers (Cooper pair transport). In our case, when we tunnel into a strictly isolated MZM (i.e., when the energy splitting $\varepsilon=0$ and $\bar{w}=0$), the particle-hole symmetry together with the requirements from the topologically non-trivial case enforce the Andreev reflection eigenvalue $T^{he}(E=0)=1$ (perfect Andreev reflection). In this case, the Fano factor will vanish in the zero voltage limit. However, Majorana fermions always come in pairs, and there are inevitable couplings and finite energy splittings between these Majorana fermions, which would bring the Fano factor back to a value of 2 in the limit of $eV\ll\varepsilon$\cite{PhysRevB.83.153415}, yielding the same result as the trivial NS junction. Therefore, it is challenging to identify the existence of MZM through measuring the Fano factor in the zero-voltage limit.
	
	By substituting Eq. (\ref{appendixCEq:1}) for the pair of MBSs, we can derive the expression for the Andreev reflection eigenvalue:
	\begin{equation}\label{appendixCEq:10}
		T^{he}(E)=\frac{\left(\Gamma_{0}+\bar{\Gamma}\right)^{2}-4\Gamma_{0}\bar{\Gamma}\sin^{2}\theta}{\left(E-\frac{\varepsilon^{2}+\Gamma_{0}\bar{\Gamma}\sin^{2}\theta}{E}\right)^{2}+\left(\Gamma_{0}+\bar{\Gamma}\right)^{2}},
	\end{equation}
	where $\Gamma_{0}=2\pi w_0^2$ and $\bar{\Gamma}=2\pi \bar{w}^2$.
	
	In the limit of $eV\gg\Gamma,\varepsilon$, we can prove that a finite energy splitting $\varepsilon$ does not affect the Fano factor, by noting the fact that the following integrals are independent of the energy shift $\lambda$.
	\begin{equation}\label{appendixCEq:12.5}
		\int_{0}^{\infty}\mathrm{d}E\,T(E),\:\int_{0}^{\infty}\mathrm{d}E\,T^{2}(E),\;\text{where}\;T(E)=\frac{\Gamma^{2}}{(E-\frac{\lambda^{2}}{E})^{2}+\Gamma^{2}}.
	\end{equation}
	Consider that $\partial_{\lambda}T(E)=4\lambda\Gamma^{2}\left[\frac{1}{(E-\lambda^{2}/E)^{2}+\Gamma^{2}}-\frac{\lambda^{2}/E^{2}}{(E-\lambda^{2}/E)^{2}+\Gamma^{2}}\right]$, the integral of $\partial_{\lambda}T(E)$ from zero to infinity vanishes since we can apply a change of variables of $E\rightarrow\lambda^2/E$ to the first term and can immediately observe the integrals of these two terms cancel exactly. For a similar reason, the integral of $T^2(E)$ is also independent of $\lambda$. Therefore, we can evaluate the integral in Eq. (\ref{appendixCEq:12.5}) at $\lambda=0$, and they turn out to be
	\begin{align}
		\int_{0}^{\infty}\mathrm{d}E\,T(E) & =\frac{\pi}{2}\Gamma\nonumber \\
		\int_{0}^{\infty}\mathrm{d}E\,T^{2}(E) & =\frac{\pi}{4}\Gamma.
	\end{align}
	With these results in hand, the Fano factor has a simple form in the high voltage limit, given by
	\begin{equation}\label{appendixCEq:13}
		F\equiv\frac{S}{2e\bar{I}}=1+\frac{2\Gamma_{0}\bar{\Gamma}(1-\cos2\theta)}{(\Gamma_{0}+\bar{\Gamma})^{2}}.
	\end{equation}
	This suggests that in the high voltage regime, the Fano factor is insensitive to the energy splitting between the two MZMs. When their wave functions overlap (i.e. $\bar{w}\sin\theta\neq 0$), the Fano factor will be greater than 1.
	
	We can also establish a connection between this Majorana toy model and the case of tunneling into a spin-polarized in-gap bound state. We start from a general tunneling Hamiltonian
	\begin{equation}
		H_{\text{tunnel}}=\frac{t}{2}\left[\psi_{T}^{\dagger}\tau_{z}\psi_{S}(j)+h.c.\right],
	\end{equation}
	where $\psi_{T}$ and $\psi_S(j)$ correspond to the Nambu spinors of the normal tip and the superconductor at site $j$, respectively, given by $\psi_{T}=\left(d_{\uparrow},d_{\downarrow},d_{\downarrow}^{\dagger},-d_{\uparrow}^{\dagger}\right)^{T}$ and $\psi_{S}(j)=\left(c_{\uparrow,j},c_{\downarrow,j},c_{\downarrow,j}^{\dagger},-c_{\uparrow,j}^{\dagger}\right)^{T}$. By projecting $\psi_{S}(j)$ onto the low energy bound states manifold,
	\begin{equation}
		\psi_{S}(j)\simeq\sum_{\pm1}\phi_{\pm1}(j)\alpha_{\pm1},\quad\alpha_{n}=\sum_{j}\phi_{n}^{\dagger}(j)\psi_{S}(j),
	\end{equation}
	where $\phi_n$ is the orthonormal eigenstate of the BdG Hamiltonian of the SC which satisfies the relation $\phi_{-n}(j)=\tau_{y}\sigma_{y}\mathcal{K}\phi_{+n}(j)$ because of the particle-hole symmetry. The fermionic operator $\alpha_n$ obeys the standard anti-commutation relations, and it holds that $\alpha_{+n}^{\dagger}=\alpha_{-n}$. Denoting the wave functions of the in-gap bound states as $\phi_{+1}(j)=\left(u_{\uparrow},u_{\downarrow},v_{\downarrow},-v_{\uparrow}\right)^{T}$ and $\phi_{-1}(j)=\tau_{y}\sigma_{y}\mathcal{K}\phi_{+1}(j)$, we can approximate the tunneling Hamiltonian as
	\begin{equation}\label{appendixCEq:16}
		H_{\text{tunnel}}=t\left[(u_{\uparrow}d_{\uparrow}^{\dagger}+u_{\downarrow}d_{\downarrow}^{\dagger})\alpha_{+1}+(v_{\uparrow}^{*}d_{\uparrow}^{\dagger}+v_{\downarrow}^{*}d_{\downarrow}^{\dagger})\alpha_{-1}\right]+h.c.
	\end{equation}
	
	We consider the case where the spins are polarized, specifically $u_{\downarrow}$ and $v_{\downarrow}$ are equal to zero. Using $\alpha_{+1}^{\dagger}=\alpha_{-1}$ , we can define two Majorana operators: $\gamma_1=\alpha_{+1}+\alpha_{+1}^{\dagger}$, $\gamma_2=-i(\alpha_{+1}-\alpha_{+1}^{\dagger})$. This allows us to simplify the tunneling Hamiltonian (\ref{appendixCEq:16}) as follows:
	\begin{equation}
		H_{\text{tunnel}}=\frac{t}{2}\left[(u_{\uparrow}+v_{\uparrow}^{*})d_{\uparrow}^{\dagger}\gamma_{1}+i(u_{\uparrow}-v_{\uparrow}^{*})d_{\uparrow}^{\dagger}\gamma_{2}\right]+h.c.
	\end{equation}
	After a gauge transformation to ensure that the coupling with Majorana 1 is real, we can directly apply Eq. (\ref{appendixCEq:13}). Consequently, in the high voltage regime, the Fano factor can be expressed as
	\begin{equation}
		F=1+\left(\frac{|u_{\uparrow}|^{2}-|v_{\uparrow}|^{2}}{|u_{\uparrow}|^{2}+|v_{\uparrow}|^{2}}\right)^{2},
	\end{equation}
	which reflects the local particle-hole asymmetry of the in-gap bound state, as stated by Eq. (\ref{sec2:eqn3}) in the main text.
	
        \section*{Appendix D: 1D Majorana Chain}
        In order to include the effect of couplings between many Majorana pairs, we consider a simple 1D Majorana tight binding model:
        \begin{equation}\label{appendixDEq:1}
        H=\sum_{j}\frac{i\varepsilon}{2}\gamma_{B,j}\gamma_{A,j}+\frac{it}{2}\gamma_{A,j+1}\gamma_{B,j}+h.c.
        \end{equation}
        Here we consider both the intra sublattice hopping $\varepsilon$ and inter sublattice hopping $t$. For $\varepsilon=t$, it is a homogeneous infinite Majorana chain, while when $\varepsilon\gg t$, it is decoupled into many independent Majorana pairs. In the case of point contact measurement, the current and shot-noise still have the form of Eq. (\ref{sec3:eqn2}) and Eq. (\ref{sec3:eqn3}). The transmission eigenvalue can be calculated by Eq. (\ref{sec4:eqn4}), which is given by $T(\omega)=-\Gamma\text{Im}G^R_{00}(\omega)$.

        We calculate its saturate Fano factor in the $eV\rightarrow\infty$ limit.
        \begin{figure}[!htbp]
        \centering
        \epsfig{figure=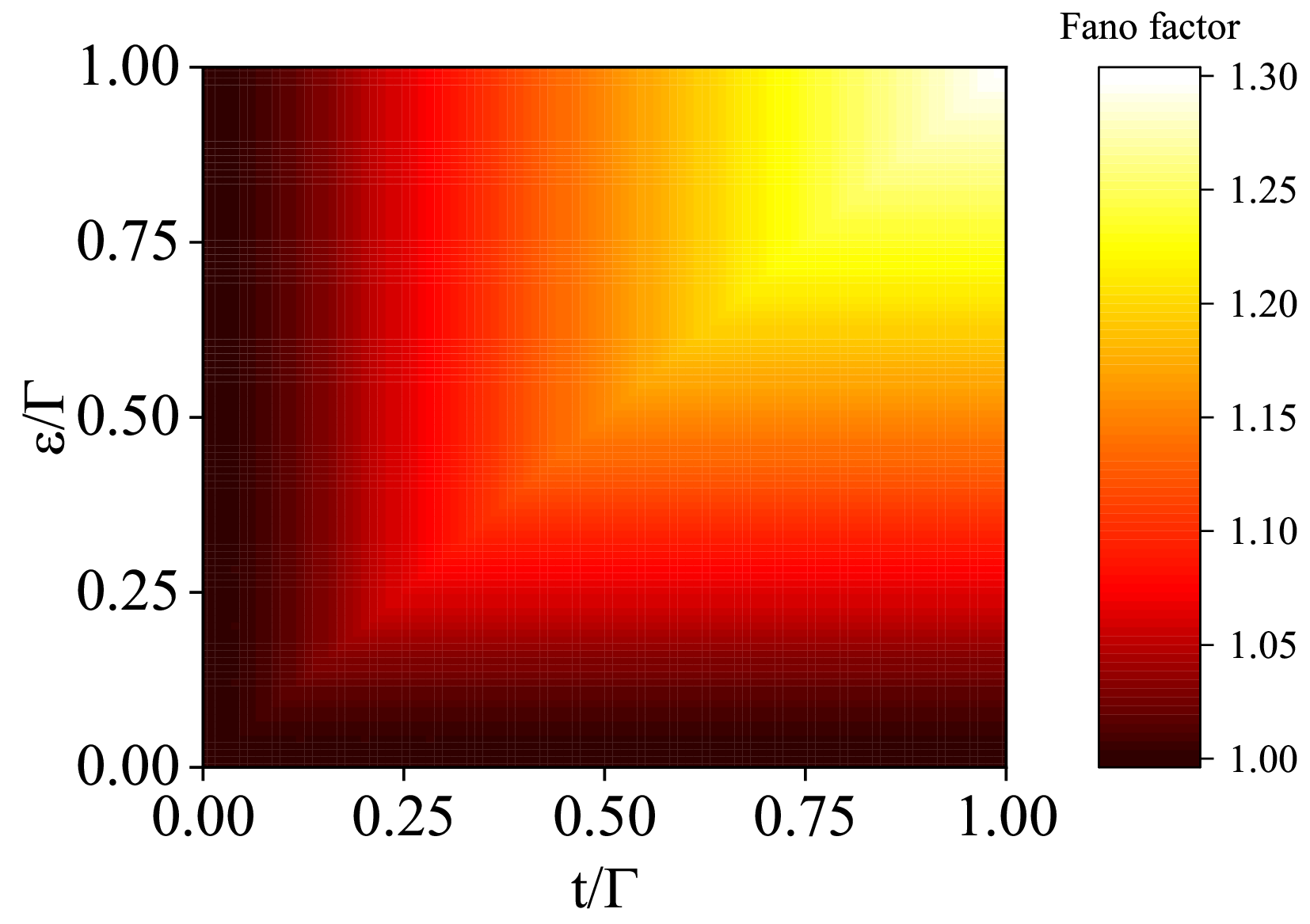, width=0.35\textwidth}
        \caption{Point contact measured Fano factors in the limit of $eV\rightarrow\infty$.}
        \label{figappendixD1}
        \end{figure}
        As demonstrated in Fig. \ref{figappendixD1}, Fano factor in the high voltage regime is not sensitive to the energy splitting $\varepsilon$ inside a MBS pair. When these Majorana pairs begin to hybridize, the Fano factor increasingly deviates from 1 as $t$ increasing. For $t\lesssim0.2\Gamma$, the observation of the Fano factors in a single pair of MBSs is still valid in the chain model.
        \begin{figure}[!htbp]
		\begin{minipage}[b]{0.47\textwidth}
			\centering
			\subfigure[]{
				\epsfig{figure=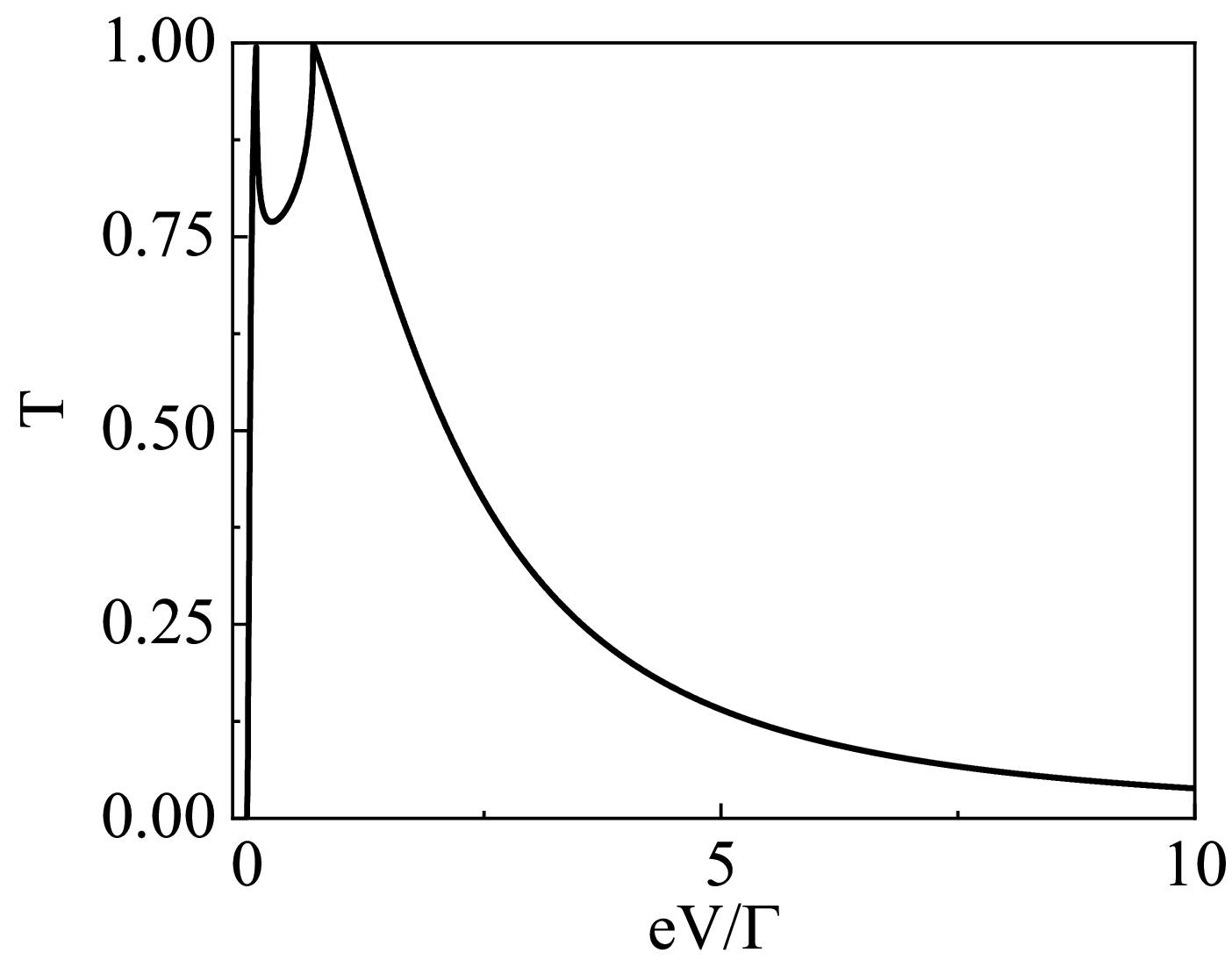, width=0.45\textwidth}\label{figappendixD2:sub1}
			}
			\subfigure[]{
				\epsfig{figure=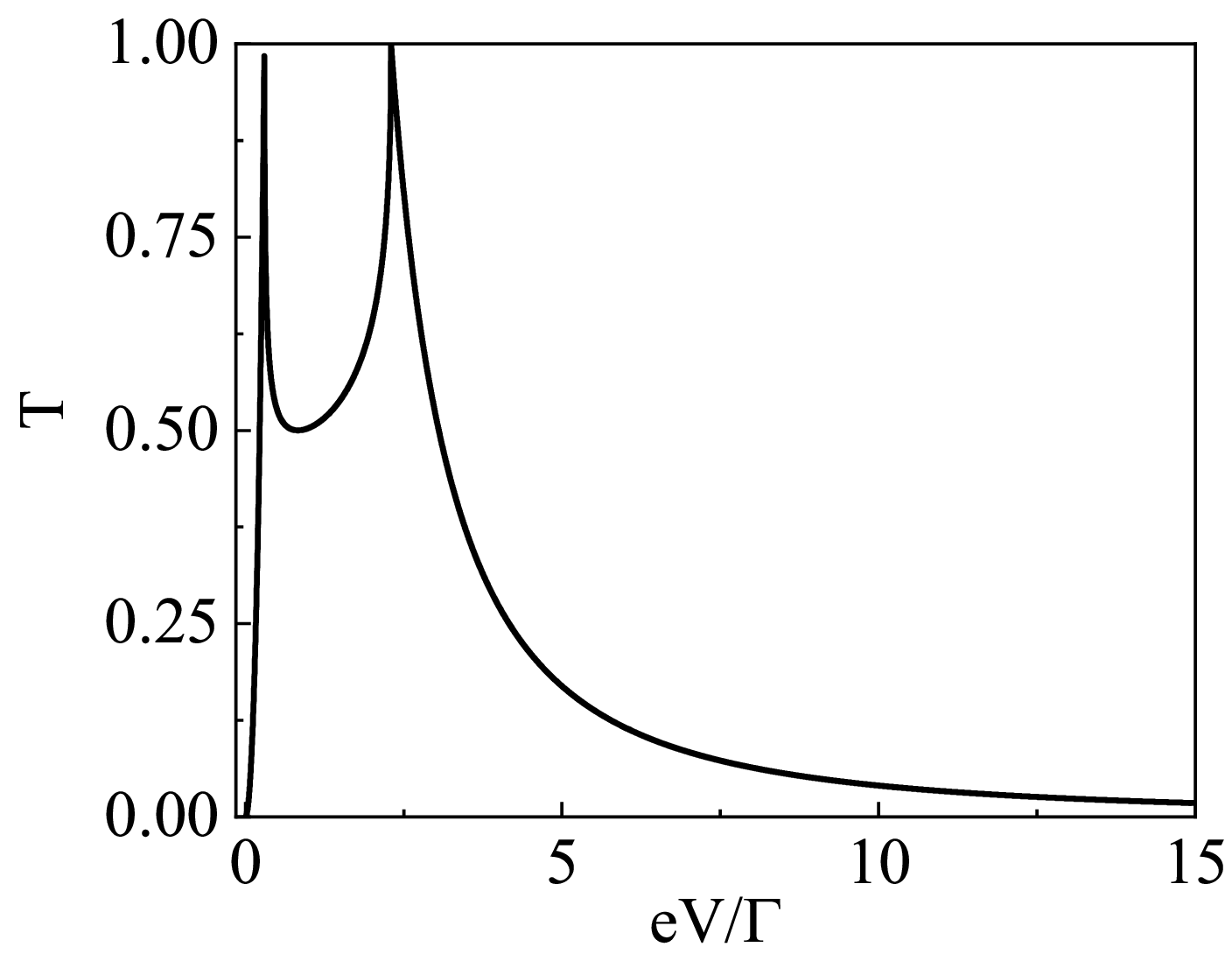, width=0.45\textwidth}\label{figappendixD2:sub2}
			}
		\end{minipage}
		
		\begin{minipage}[b]{0.5\textwidth}
			\centering
			\subfigure[]{
				\epsfig{figure=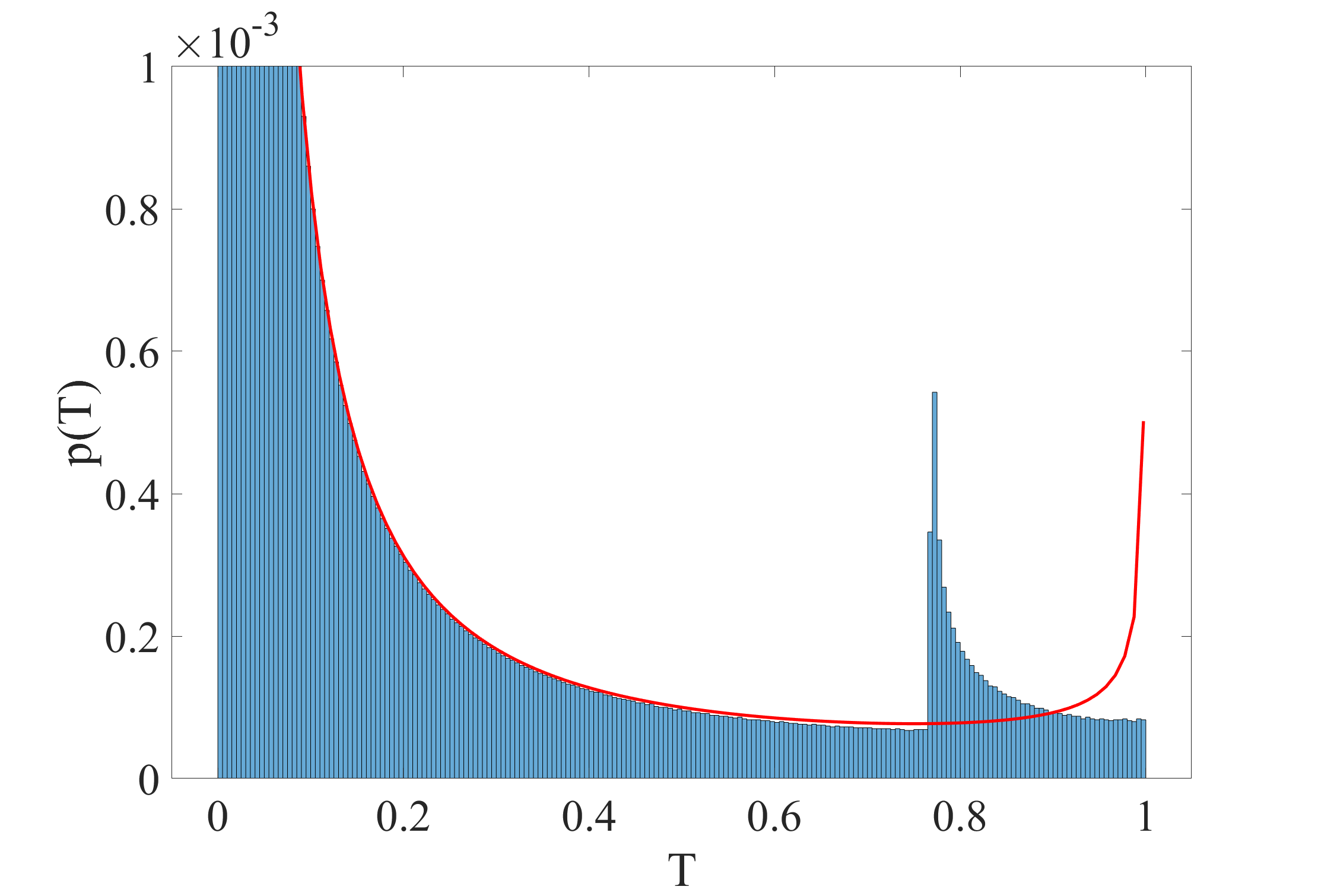, width=0.45\textwidth}\label{figappendixD2:sub3}
			}
			\subfigure[]{
				\epsfig{figure=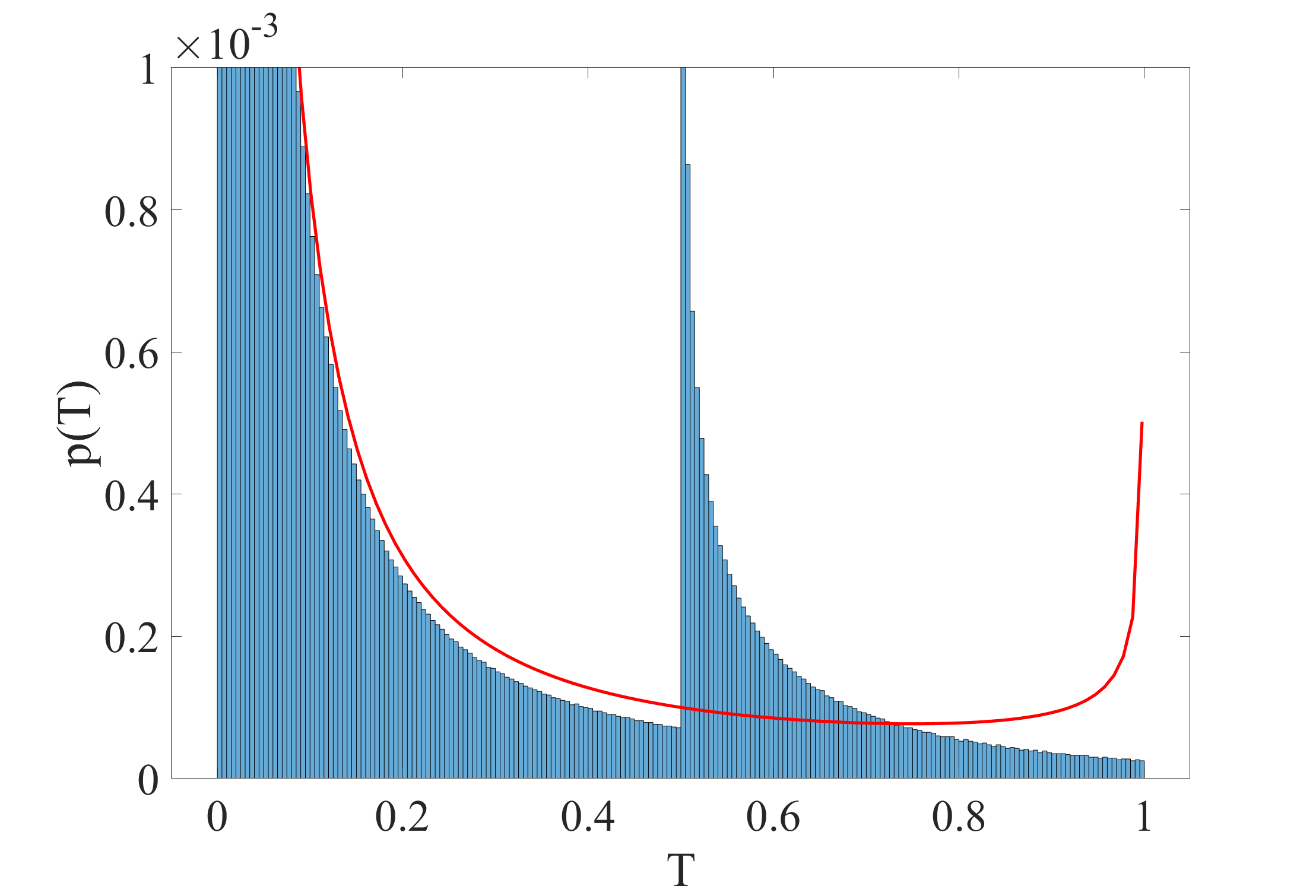, width=0.45\textwidth}\label{figappendixD2:sub4}
			}
		\end{minipage}
		\caption{Transmission eigenvalues $T(\omega)$ at different $\varepsilon,\,t$ parameters and their corresponding distributions $p(T)$. (a) $\varepsilon=0.2\Gamma,\,t=0.15\Gamma$ (b) $\varepsilon=0.65\Gamma,\,t=0.5\Gamma$. The red line indicates the distribution $p(T)\propto\frac{1}{T^{3/2}\sqrt{1-T}}$.}
		\label{figappendixD2}
	\end{figure}
 
        We further choose several parameters and calculate the transmission eigenvalue $T(\omega)$ and the corresponding distribution $p(T)$. Fig. \ref{figappendixD2:sub1} and Fig. \ref{figappendixD2:sub2} show the transmission eigenvalue $T(\omega)$ for two sets of parameters, and the corresponding Fano factors measured in the high voltage limit are $F=1.03$ and $F=1.15$ respectively. However, as demonstrated in Fig. \ref{figappendixD2:sub3} and Fig. \ref{figappendixD2:sub4}, the distribution $p(T)$ deviates from the universal distribution mentioned in the main text in both cases. It turns out that $p(T)$ may be hard to characterize $F(\infty)$ quantitatively in a system consisting of many Majorana pairs.
        
	\section*{Appendix E: YSR bound states}
\begin{figure}[!htbp]
		\centering
		\subfigure[]{
			\epsfig{figure=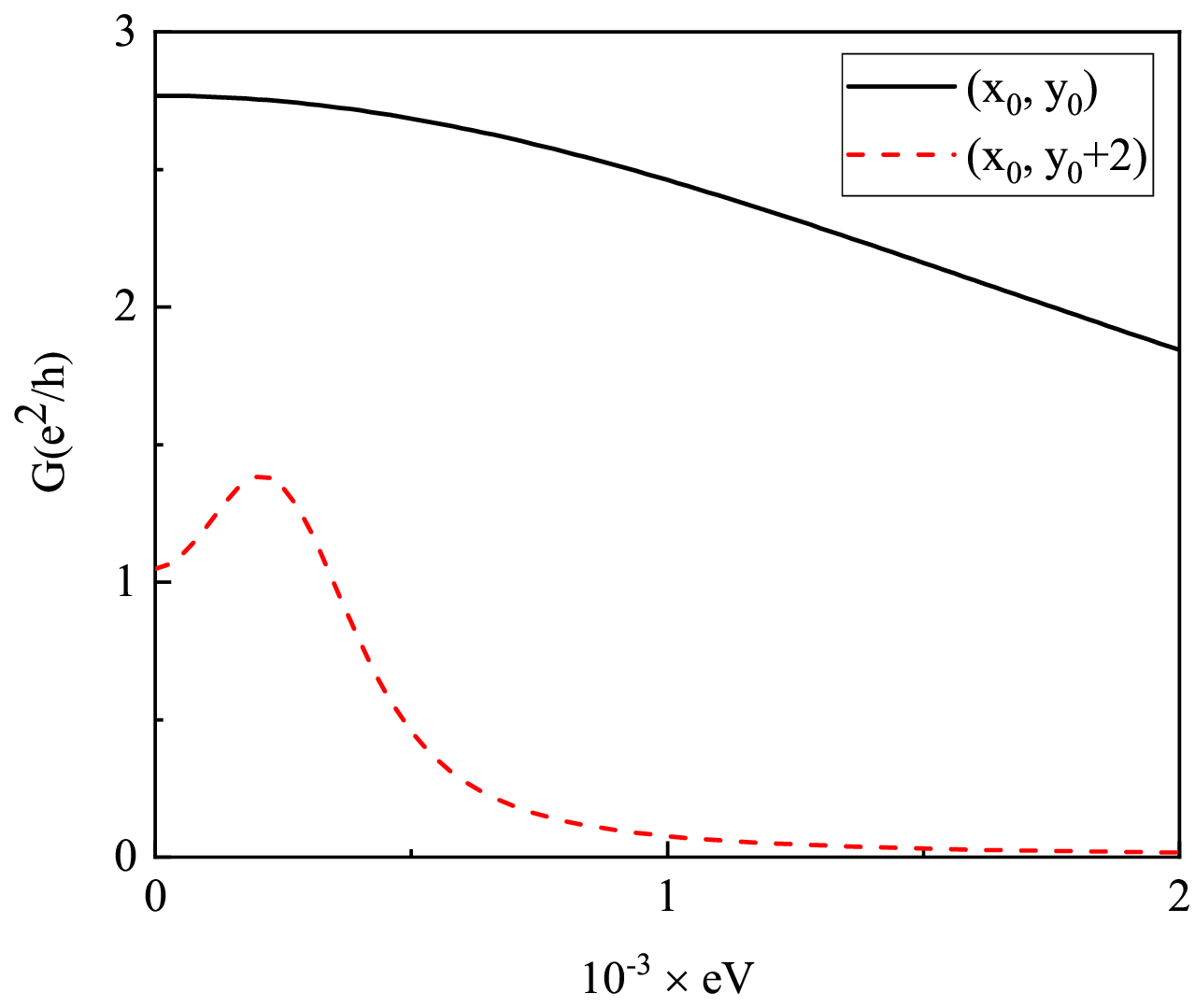, width=0.23\textwidth} \label{fig:Conductance_of_SingleYSR}
		}
		\subfigure[]{
			\includegraphics[width=0.23\textwidth]{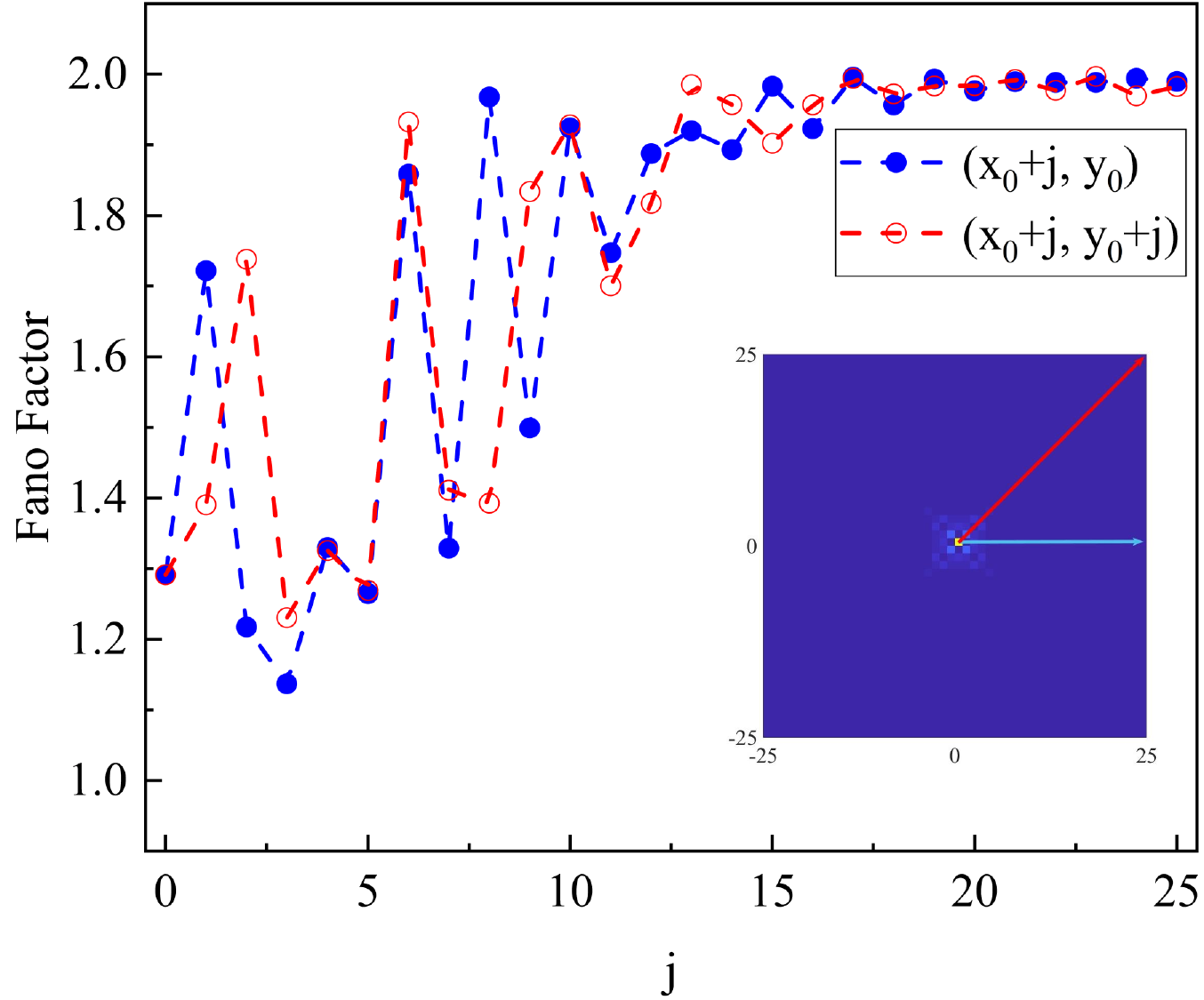}\label{fig:Single YSR FanoFactors}
		}
		\caption{(a) Differential conductance of a YSR bound state. (b) Spatially resolved Fano factors in the 2D plane. The coordinate $(x_0, y_0)$ represents the location of the magnetic impurity. The simulation was performed using a 2D square lattice tight-binding model with nearest-neighbor hopping strength $t_1=-1.5$ and a chemical potential $\mu=3.5$, measured from the bottom of the band. A pairing field of $\Delta=0.4$ and an exchange strength of $JS_z=3.45$ were set to induce a deep YSR bound state. In the tunneling simulations, an energy width of $\Gamma=0.1\Delta$, a relaxation parameter of $\eta=1\times10^{-5}$ and an inverse temperature of $\beta=8000/\Delta$ were used. For the Fano factor simulation (b), a fixed voltage bias of $V_{\text{bias}}=0.2$ was applied.}
		\label{figappendixE1}
	\end{figure}

	We also analyze the case of Yu-Shiba-Rusinov (YSR) states in our study. For a single magnetic impurity, we consider the BdG Hamiltonian,
	\begin{equation}
		H=H_0+H_{imp}=\xi_{k}\tau_{3}\sigma_{0}+\Delta\tau_{1}\sigma_{0}+JS_{z}\tau_{0}\sigma_{3}\delta(\bm{r}).
	\end{equation}
	Here, we assume the impurity potential is purely local, with the impurity spin $\bm{S}$ pointing along the $z$ direction. In the wide band limit, the energy of the bound state is given by\cite{Rusinov1969,RevModPhys.78.373} $\varepsilon=\Delta\cos(2\delta_0)$, where $\tan\delta_0 = \pi\nu_0JS_z$ and $\nu_0$ represents the normal state density of states at the Fermi level. The bound state wave functions at the impurity location are\cite{PhysRevB.88.155420,doi:10.1038/nphys3508}:
	\begin{equation}
		\phi_{+}(0)=\frac{1}{\sqrt{\mathcal{N}}}\begin{pmatrix}1\\
			0\\
			-1\\
			0
		\end{pmatrix},\qquad\phi_{-}(0)=\frac{1}{\sqrt{\mathcal{N}}}\begin{pmatrix}0\\
			1\\
			0\\
			1
		\end{pmatrix}.
	\end{equation}
	The behavior of the wave function away from the impurity depends strongly on the dimensionality of the system. In three dimensions (3D), the wave function decays as $\frac{1}{r}\exp\left(-r\sqrt{\Delta^2-\varepsilon^2}/\hbar v_F\right)$. However, in two dimensions (2D), it decays much slower, going as\cite{doi:10.1038/nphys3508} $\frac{1}{\sqrt{r}}\exp\left(-r\sqrt{\Delta^2-\varepsilon^2}/\hbar v_F\right)$. Here, we focus on the 2D case, where the wave function away from the impurity can be expressed as
	\begin{equation}
		\phi_{+}(r)=\frac{1}{\sqrt{\mathcal{N}\pi k_Fr}}\begin{pmatrix}\sin(k_{F}r-\frac{\pi}{4}+\delta_{0})\\
			0\\
			-\sin(k_{F}r-\frac{\pi}{4}-\delta_{0})\\
			0
		\end{pmatrix}\exp\left(-\frac{\sqrt{\Delta^{2}-\varepsilon^{2}}}{\hbar v_{F}}r\right),
	\end{equation}
	where $\mathcal{N}$ is a normalization factor and $\phi_{-}(r)=\tau_y\sigma_y\mathcal{K}\phi_{+}(r)$. Clearly, the YSR bound state exhibits spatially oscillating electron-hole asymmetry, which can potentially be characterized through Fano factor tomography. In the case of deep YSR states, corresponding to a dephasing $2\delta_0\rightarrow\pm\pi/2$, the electron and hole YSR density of states exhibit antiphase behavior far from the impurity. This antiphase behavior leads to a strong spatial oscillation of the Fano factor as outlined in Eq. (\ref{sec2:eqn3}). The Appendix Figure \ref{figappendixE1} demonstrates the differential conductance and the spatially resolved Fano factors of a deep YSR state. The deep YSR state exhibits a ZBCP signature, although it is not quantized. The Fano factor tomography of the deep YSR state strongly oscillates near the magnetic impurity and saturates at a value of 2 due to the dominant contribution of Andreev reflections from bulk states.
	
\begin{figure}[!htbp]
	 	\begin{minipage}[b]{0.5\textwidth}
	 		\centering
	 		\subfigure[]{
	 			\epsfig{figure=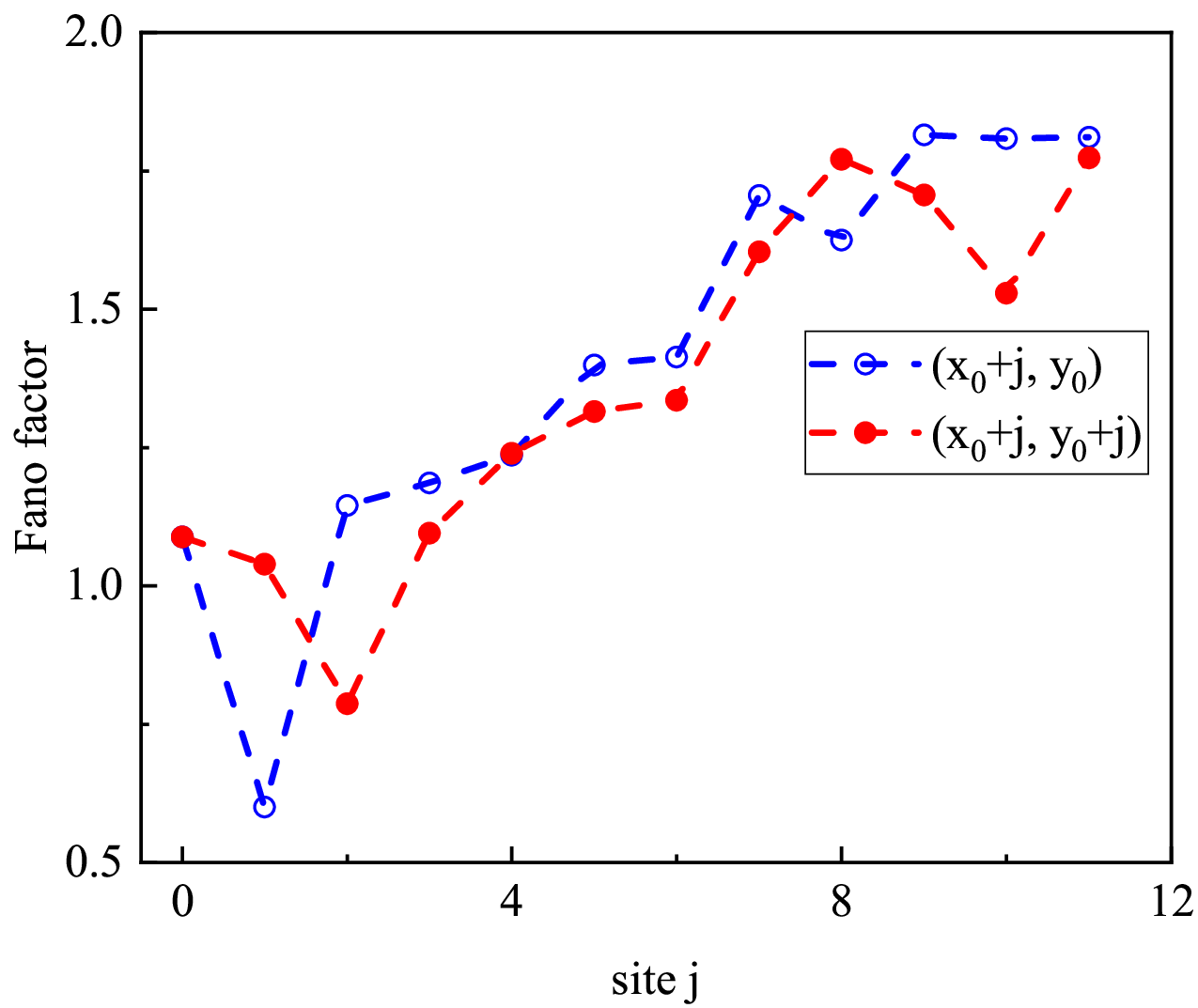, width=0.45\textwidth} \label{fig:F(j)YSRLattice_N_23,1e-5}
	 		}
	 		\subfigure[]{
	 			\epsfig{figure=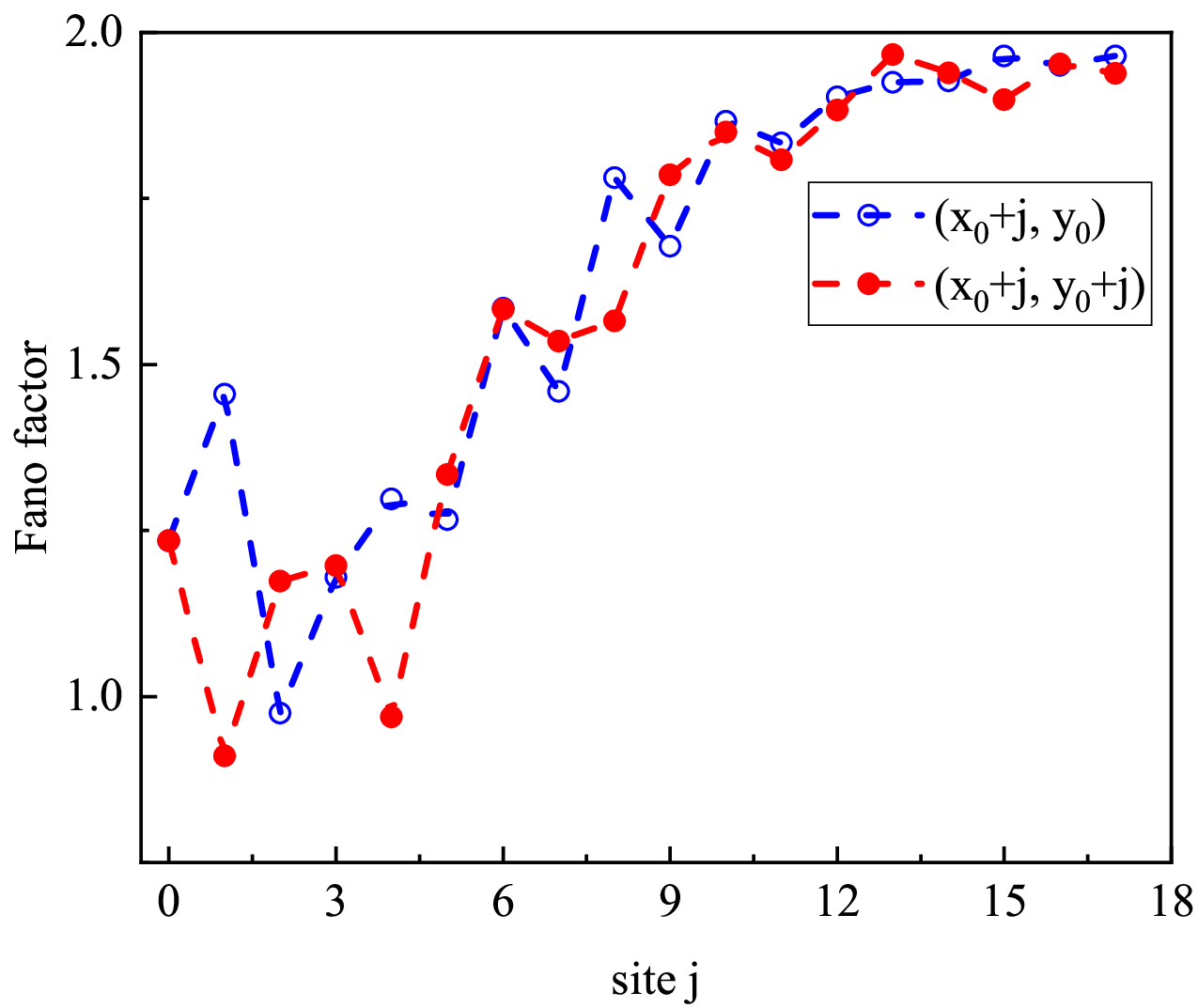, width=0.45\textwidth} \label{fig:F(j)YSRLattice_N_35,1e-5}
	 		}
	 	\end{minipage}
	 	
	 	\begin{minipage}[b]{0.5\textwidth}
	 		\centering
	 		\subfigure[]{
	 			\epsfig{figure=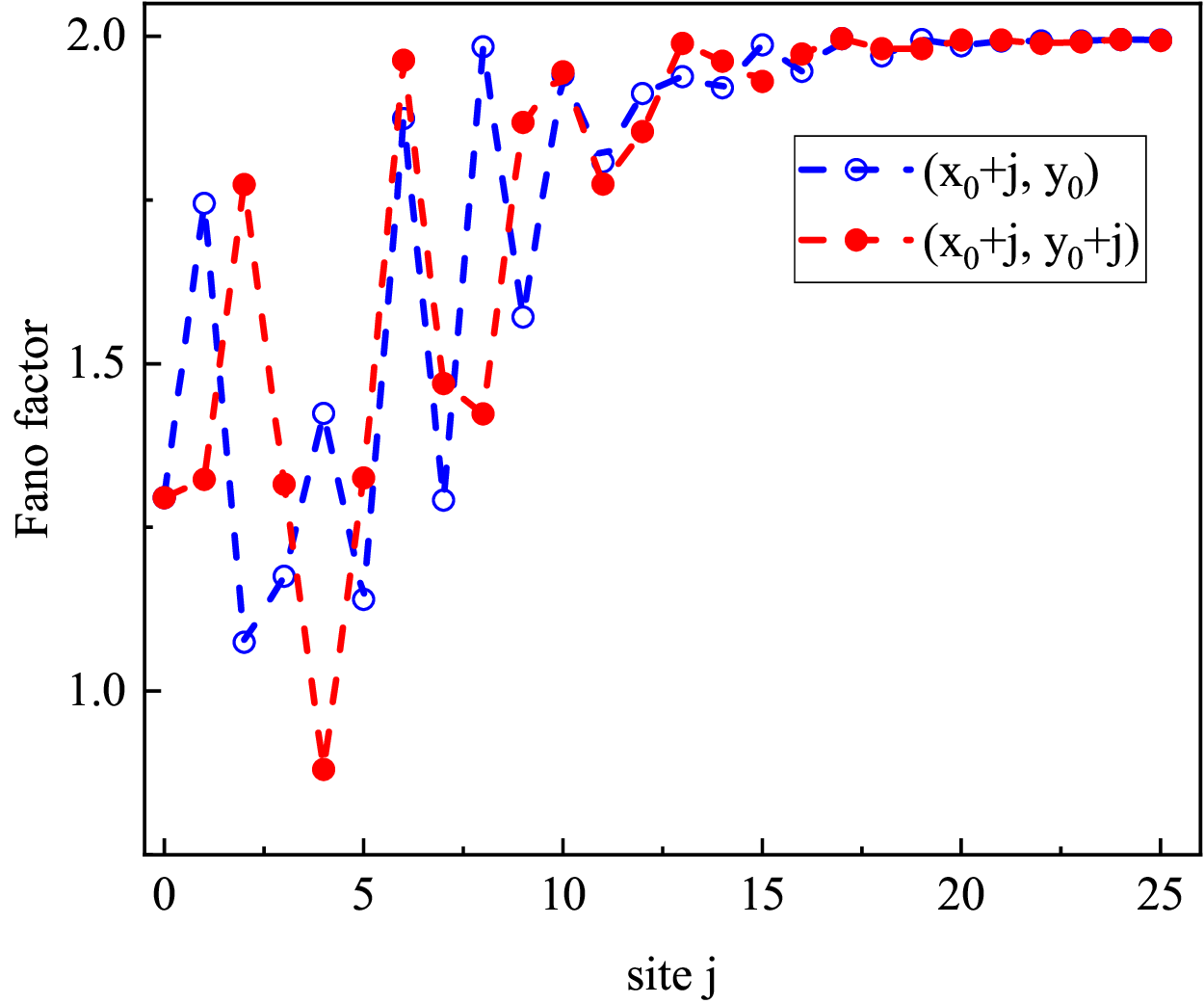, width=0.45\textwidth} \label{fig:F(j)YSRLattice_N_51,1e-5}
	 		}
	 		\subfigure[]{
	 			\epsfig{figure=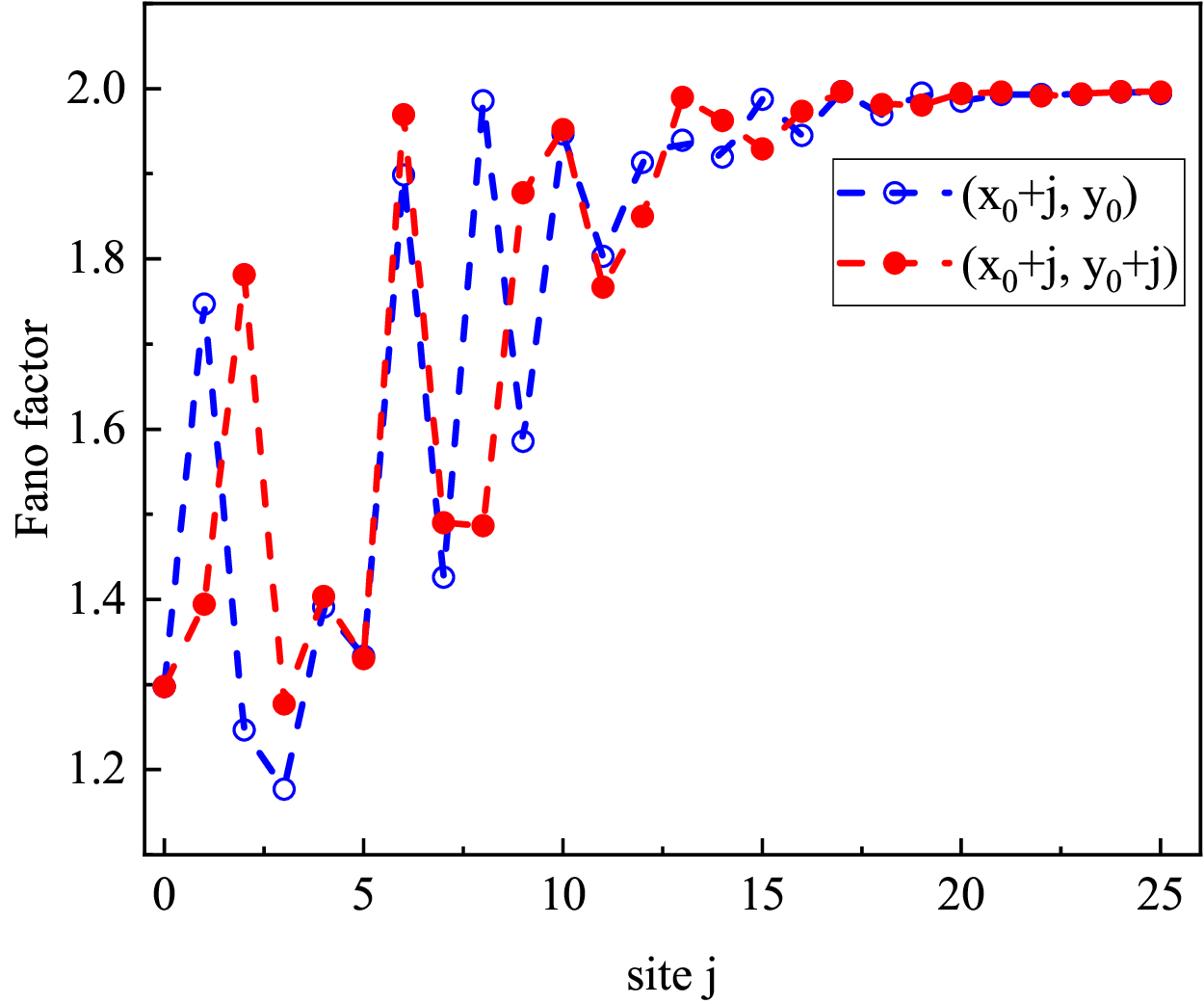, width=0.45\textwidth} \label{fig:F(j)YSRLattice_N_75,1e-5}
	 		}
	 	\end{minipage}
	 	\caption{Fano factor tomography of magnetic impurity lattices with different distances $d$ between impurities is presented. The distances are specified as: (a) $d=23a$, (b) $d=35a$, (c) $d=51a$, (d) $d=75a$, where $a$ represents the lattice constant of the tight-binding model introduced earlier. The coordinates of one of the magnetic impurities are denoted as $(x_0, y_0)$. In the simulations, we maintain a fixed applied voltage of $V_{\text{bias}}=0.25$, while all other parameters remain the same as in Figure \ref{figappendixE1}.}
	 	\label{figappendixE2}
	 \end{figure}

	We also perform Fano factor tomography calculations in the presence of multiple magnetic impurities. To simplify the calculations, we assume that the magnetic impurities have the same exchange strength $JS_z$ and are periodically arranged into a square lattice. The numerical results demonstrate that as long as the impurity lattice is not too dense, the Fano factor tomography remains effective in capturing the signature of spatially oscillating electron-hole asymmetry in the YSR bound states, as shown in Fig. \ref{figappendixE2}. However, as the impurities become denser, the Fano factors near a magnetic impurity decrease. This reduction can be attributed to the increased occurrence of single-electron tunneling processes. Specifically, as the impurities get closer, the overlaps between their respective bound states become more pronounced, leading to an increase in the effective single-electron tunneling channels and contributing to the single-electron tunneling current.
	
\clearpage
\nocite{*}
\bibliography{references}

\begin{thebibliography}{50}
\bibitem [{rem()}]{remark1}%
  \BibitemOpen
  \href@noop {} {}\bibinfo {note} {The derivation can be seen in the Appendix
  C.}\BibitemShut {Stop}%
\end{thebibliography}
\end{document}